\documentclass[
amssymb,preprint,preprintnumbers,nofootinbib,superscriptaddress]{revtex4}
\usepackage{amsmath}
\usepackage{graphicx}
\graphicspath{{Figures/}}
\usepackage{latexsym}
\usepackage{amsfonts}
\usepackage{url,hyperref}
\usepackage{bm}
\usepackage{textcomp}
\usepackage{color}
\usepackage{bbm}
\usepackage{slashed}
\usepackage{caption}
\usepackage{epstopdf}
\usepackage{subcaption}
\usepackage{placeins}
\usepackage{color}

\begin{document}

\title{Effective thermodynamical system of \\ Schwarzschild-de Sitter black holes from R\'{e}nyi statistics}

\author{Ratchaphat Nakarachinda \footnote{Email: tahpahctar\_net@hotmail.com}}
\affiliation{The Institute for Fundamental Study, Naresuan University, Phitsanulok 65000, Thailand}

\author{Ekapong Hirunsirisawat \footnote{Email: ekapong.hir@mail.kmutt.ac.th}}
\affiliation{Theoretical and Computational Physics Group, Theoretical and Computational Science Center (TaCS), Faculty of Science, King Mongkut’s University of Technology Thonburi, 126 Pracha Uthit Road, Bang Mod, Thung Khru, Bangkok 10140, Thailand}
\affiliation{Learning Institute, King Mongkut’s University of Technology Thonburi, 126 Pracha Uthit Rd., Bang Mod, Thung Khru, Bangkok 10140, Thailand}

\author{Lunchakorn Tannukij \footnote{Email: l\_tannukij@hotmail.com}}
\affiliation{Theoretical and Computational Physics Group, Theoretical and Computational Science Center (TaCS), Faculty of Science, King Mongkut’s University of Technology Thonburi, 126 Pracha Uthit Road, Bang Mod, Thung Khru, Bangkok 10140, Thailand}
\affiliation{KOSEN-KMITL, King Mongkut Institute of Technology Ladkrabang, 1 Chalong Krung 1 Alley, Lat Krabang, Bangkok 10520, Thailand}

\author{Pitayuth Wongjun \footnote{Email: pitbaa@gmail.com}}
\affiliation{The Institute for Fundamental Study, Naresuan University, Phitsanulok 65000, Thailand}
\affiliation{Thailand Center of Excellence in Physics, Ministry of Higher Education, Science, Research and Innovation, 328 Si Ayutthaya Road, Bangkok 10400, Thailand}

\begin{abstract}
It has been known that the Schwarzschild-de Sitter (Sch-dS) black hole may not be in thermal equilibrium and also be found to be thermodynamically unstable in the standard black hole thermodynamics. In the present work, we investigate the possibility to realize the thermodynamical stability of the Sch-dS black hole as an effective system by using the R\'{e}nyi statistics, which includes the nonextensive nature of black holes. Our results indicate that the nonextensivity allows the black hole to be thermodynamically stable, which gives rise to the lower bound on the nonextensive parameter. By comparing the results to ones in the separated system approach, we find that the effective temperature is always smaller than the black hole horizon temperature and the thermodynamically stable black hole in the effective approach is always larger than the one in the separated approach at a certain temperature. There exists only the zeroth-order phase transition from the hot gas phase to the black hole phase for the effective system, while it is possible to have transitions of both zeroth order and first order for the separated system.

\end{abstract}
\maketitle{}


\section{Introduction}\label{sect: intro}
The discovery that a black hole behaves as a thermal object has led to a clue in linking general relativity, thermodynamics, and quantum field theory~\cite{Hawking:1974sw,Bekenstein:1973ur,Bardeen:1973gs}. Rooted from the idea that the Hawking radiation process has a quantum origin, black hole thermodynamics has guided us to research questions relating to quantum information theory and some quantum aspects of gravity~\cite{Jacobson:1993vj,Youm:1997hw,Ashtekar:1997yu,Carlip:1999db,Solodukhin:2011gn,Bianchi:2012br}.  Interestingly, the unitarity of black hole evaporation and the quantum origin of black hole entropy have are still active research topics nowadays.   Moreover, the studies of black holes in anti--de Sitter space (AdS) allow us to observe fascinating thermodynamical behaviors of black holes.  One of these is the Hawking-Page phase transition which has a natural interpretation in AdS/CFT~\cite{Maldacena:1997re} as a deconfinement transition in the boundary CFT~\cite{Witten:1998zw}.  

The phase structure of the AdS black holes has been vastly explored in extended phase space.  In this framework, the cosmological constant, in particular, $\left|\Lambda\right|=-\Lambda$, can  be treated as a thermodynamical variable acting as pressure~\cite{Kastor:2009wy}.  A variety of phenomena from these studies indicate that its workings are similar way to thermodynamical phenomena appearing in everyday life such as the liquid-gas phase transition. Unexpectedly, some of these correspond with some phenomena of exotic quantum matter.  These give a nontrivial perspective on the phenomenological thermodynamical behaviors of the dual field theory via the gauge-gravity correspondence~\cite{Karch:2015rpa,Sinamuli:2017rhp,Wen:2007vy}. It is important to emphasize here that extended phase space can be applied not only in asymptotically AdS black holes, but also in asymptotically de Sitter (dS) black holes.

The dS space should attract a high level of research interest, since its corresponding cosmological model is potentially consistent with the observational results supporting the late-time accelerating expansion of the Universe~\cite{Aghanim:2018eyx}.  In spite of this, the thermodynamical phenomena in asymptotically dS black holes remains slightly explored, while those in asymptotically AdS black holes have been substantially discovered.  Theoretically, thermodynamical studies of Schwarzschild--de Sitter (Sch-dS) black holes have historically encountered a number of difficulties in considerations.  Among plenty of issues obstructing progress, the obscurities in dealing with the notion of mass and with thermodynamical consideration of a multihorizon system are two main problems.  The former is due to the absence of the globally timelike Killing vector field, which prevents one to have a well-defined asymptotic mass~\cite{Bousso:2002fq,Clarkson:2005qx,Clarkson:2003kt,Ashtekar:2014zfa}. The latter results from the fact that the black hole  and cosmic horizons are generically at different temperatures, such that the Sch-dS black hole as a multihorizon system is out of equilibrium. 
 
There are several solutions proposed in dealing with the issue of different temperatures in multihorizon systems. One straightforward approach is to consider each horizon as a separated thermodynamical system, characterized by its own thermodynamical behavior. This approach can give some perspectives on the thermodynamical behaviors. For example, the sign of a phase transition of the whole Sch-dS black hole may be obtained from the considerations of phase transitions occurring in separated thermodynamical subsystems~\cite{Kubiznak:2015bya}.
Since the separated system approach cannot give a complete understanding of the phase structure of the entire system, the effective temperature approach is proposed, where a single temperature is assigned to the entire spacetime.  Considering the thermodynamical description from which an observer is located in between the black hole horizon and cosmic horizon, the observer experiences an effective temperature $T_\text{eff}$, which can be derived through a postulated thermodynamical first law~\cite{Urano:2009xn,Ma:2013aqa,Zhao:2014raa,Zhang:2014jfa,Ma:2014hna,Guo:2015waa,Guo:2016eie}. It is worthwhile to remark that the effective temperature can also be obtained through a Euclidean path integral approach with fixing the temperature at the boundary of the cavity between the event and  cosmic horizons~\cite{Simovic:2020dke}.

In the literature, the mass parameter can be interpreted as either the internal energy $E$ or the enthalpy $H$ in the effective approach, each of which gives rise to a different effective thermodynamical description and different $T_\text{eff}$ (see \cite{Kubiznak:2016qmn} for a review). Note that, for simplicity, we will discuss only spherically symmetric dS black holes in this work. Treating the mass $M$ as the internal energy $E$ with the first law of the form $\delta E = T_\text{eff} \delta S - P_\text{eff} \delta V$, the total entropy $S$ of this effective system is given by the sum of the black hole event horizon and cosmic horizon entropies, while the thermodynamical volume of the effective system equals the volume of the observable Universe.  These take the form~\cite{Kastor:1992nn,Bhattacharya:2015mja}
\begin{eqnarray}
	S=S_b+S_c, \qquad V=V_c-V_b, \qquad E=M.
\end{eqnarray} 
Unfortunately, this version of the effective temperature approach encounters some problems causing an unclear physical picture for more complicated black hole spacetimes~\cite{Kubiznak:2016qmn}. Additionally, neither the effective temperature nor the effective pressure are apparently positive.   As another version of the effective approach, the mass $M$ is on the other hand treated as a gravitational enthalpy with the redefinition of the pressure taking the form $\mathcal{P}=-P\propto\Lambda$.  The effective first law is, thus,  $\delta H = T_\text{eff} \delta S + V_\text{eff} \delta \mathcal{P}$.  Note that the thermodynamical state variables, other than pressure, relate to the geometrical variables through~\cite{Li:2016zca}
\begin{eqnarray} \label{eff_enthal}
	S=S_b+S_c, \qquad V_\text{eff}= \left( \frac{\partial H}{\partial \mathcal{P}}\right)_S,  \qquad H=-M.
\end{eqnarray}
Remark that the effective volume $V_\text{eff}$ is no longer $V_c-V_b$. Rather, it is the conjugate of the pressure which can be derived from the first law as shown in Eq.~\eqref{eff_enthal}.  Accordingly, we have the form of the effective temperature as
\begin{eqnarray}\label{Teff}
	T_\text{eff} = \left( \frac{1}{T_c}-\frac{1}{T_b}\right)^{-1}.
\end{eqnarray}
However, the effective temperature of this form can render unphysical properties.  For example, considering Eq.~\eqref{Teff}, the value of $T_\text{eff}$ blows up to infinity at the Nariai or lukewarm limit, i.e., $T_b=T_c$.  Moreover, for the charged black hole case, it encounters an infinite jump when $T_b=T_c$ and becomes negative for a range of black hole horizons.

Interestingly, one of the solutions of this problem can be obtained by identifying the effective entropy as
\begin{eqnarray} \label{S_minus}
	S=S_c-S_b, 
\end{eqnarray}
as an \textit{ad hoc} condition, such that one obtains~\cite{Kubiznak:2016qmn,Kanti2017}
\begin{eqnarray} \label{TeffVeff}
	T_\text{eff}=\left( \frac{1}{T_c}+\frac{1}{T_b}\right)^{-1}, \qquad V_\text{eff}=T_\text{eff} \left(\frac{V_c}{T_c}+\frac{V_b}{T_b}\right),  
	\end{eqnarray} 
where $T_\text{eff}$ and $V_\text{eff}$ are apparently positive here. As will be seen in the present paper, we use the expression of total entropy 
\begin{eqnarray} \label{S_plus}
	S=S_b+S_c, 
\end{eqnarray}
rather than Eq.~\eqref{S_minus}. Despite this, the change of total entropy in Eq.~\eqref{S_plus} obtained from the consideration of heat flow into the spacetime region between these two horizons where the  observer sits, can give the terms associated with $S_b$ and $S_c$ of opposite sign when we identify the direction of heat flow at the cosmic horizon to be opposite to the one at the black hole horizon. In other words, we live between these two horizons such that the directions of heat flows with respect to the changes of the horizon radii would be opposite.  As a consequence of this identification, the change of total entropy turns out to take the form  
\begin{eqnarray} \label{heat_S_minus}
	\text{d}S &= &\text{d} S_b-\text{d} S_c, 
\end{eqnarray}
where we have used the fact that the heat flowing from outside the cosmic horizon to inside corresponds to the decrease of entropy satisfying the first law of the cosmic horizon system [see Appendix~\ref{app: eff quan deriv} for a further discussion on the signs of Eq.~\eqref{heat_S_minus}]. Fortunately, the resulting effective temperature $T_\text{eff}$ and volume $V_\text{eff}$ from our setting, using Eq.~\eqref{S_plus}, are identical to the above expressions as shown in Eq.~\eqref{TeffVeff} obtained from using Eq.~\eqref{S_minus}. Moreover, in the case of $M$ being the internal energy,  the effective temperature will still be well behaved by using this criteria.

As seen in both versions of the effective temperature approach, the entropy of the effective system is simply the sum of $S_b$ and $S_c$.  Nevertheless, as often said, the whole is not a sum of its parts.  It has been argued by He, Ma and Zhao \cite{He:2018zrx} that, considering the dS black holes, the entropy of the effective system may not be simple like that. In other words, even though there is still no consensus on how one can describe mathematically the microstates of a black hole event horizon and cosmic horizon, the total number of microscopic states of this hypothetical correlated system is probably not the product of those of two horizons. Thus, the total entropy is proposed to be in the form 
\begin{eqnarray}
	S=S_b+S_c+S_{ex},
\end{eqnarray}
where $S_{ex}$ is the extra entropy term responsible to the correlations between these two horizons.    

Whereas the work of Ref.~\cite{He:2018zrx} uses the bottom-up model in which $S_\text{ex}$ is an arbitrary function to be determined through the consistency tests in a variety of phenomenological aspects, a question may arise how one can formulate naturally the thermodynamical description associated with the long-range correlations between these two subsystems of the multihorizon system.  In the present paper,  we make an attempt to investigate the entire Sch-dS black hole thermodynamical system as a correlated system through a top-down model based on nonextensive thermodynamics, instead of the conventional Gibbs-Boltzmann (GB) thermodynamics.
 
As evident from the area law, the black hole system has been argued to be a nonextensive system in its own right~\cite{Tsallis:2012js}. Having considered black hole systems using the Tsallis and R\'enyi statistics can provide a number of new perspectives on its thermodynamical behaviors~\cite{Tsallis:2012js,Czinner:2015eyk,Czinner:2017tjq,Promsiri:2020jga,Tannukij:2020,Samart:2020klx,Promsiri2021}. Taking into account the effect of nonextensivity can not give only the thermodynamical stability in some regions of parameter, it also allows a phase transition occurring in the case of black holes in asymptotically flat spacetime~\cite{Czinner:2015eyk,Czinner:2017tjq,Promsiri:2020jga,Promsiri2021}. Interestingly, Tannukij {\it et al.} \cite{Tannukij:2020} have demonstrated that the separated black hole system of the dS black hole spacetime can be in thermodynamical stability when the deviation from the conventional GB thermodynamics is large enough. These features resulted from the fact that the nonextensive thermodynamics is the area that has included the effect of the correlations, or perhaps long-range interactions, between the environment spacetime and the black hole event horizon. This causes the modified thermodynamical description depending on the nonextensivity parameter. To achieve the nonextensivity approach, we need to relax the additive composition rule, one of the Shannon-Khinchin axiomatic definitions of the entropy function, to a weaker nonadditive composition rule~\cite{Abe2001}. Considering two systems with the correlations between them, the nonextensive Tsallis entropy, for instance, follows the composition rule~\cite{Tsallis:1987eu} 
\begin{eqnarray} \label{Tsallis com}
	S_\text{T}^{12}=S_\text{T}^1+ S_\text{T}^2 +\lambda S_\text{T}^1 S_\text{T}^2, 	
\end{eqnarray}
where $S_\text{T}^{12}$ is the Tsallis entropy of the entire system, $S_\text{T}^1$ and $S_\text{T}^2$ are the Tsallis entropies of the two separated systems, and $\lambda$ is the nonextensive parameter; its value vanishes when the system begins following the GB statistics. Remark that Eq.~\eqref{Tsallis com} satisfies the Abe nonadditive entropy composition rule~\cite{Abe2001}. Requiring zeroth law compatibility, the issue of the unclear definition of the empirical temperature can be solved by transforming the Tsallis entropy into the R\'enyi entropy~\cite{Renyi:1959,Renyi:1961, Biro2011}:
\begin{eqnarray} \label{Renyi Log}
	S_\text{R}^{12}=\frac{1}{\lambda} \ln \left[ 1+ \lambda S_\text{T}^{12}\right]. 	
\end{eqnarray}
Manifestly, it is the logarithmic form of the R\'enyi entropy that makes the composition rule of the Tsallis entropy turn out to be additive, despite the presence of its nonextensive nature.  This can be  shown easily by substituting Eq.~\eqref{Tsallis com} into Eq.~\eqref{Renyi Log}, then we obtain 
\begin{eqnarray} \label{Renyi com}
	S_\text{R}^{12}
	&=&\frac{1}{\lambda} \ln \left[ 1+\lambda S_\text{T}^1 +\lambda S_\text{T}^2 + \lambda^2 S_\text{T}^1 S_\text{T}^2\right]\nonumber\\
	&=&\frac{1}{\lambda} \ln  \left[ \left(1+\lambda S_\text{T}^1 \right) \left( 1+ \lambda S_\text{T}^2 \right)\right]\nonumber\\
	&=&\frac{1}{\lambda} \ln  \left[ 1+\lambda S_\text{T}^1 \right]+\frac{1}{\lambda} \left[ 1+ \lambda S_\text{T}^2\right]\nonumber\\
	&=&S_\text{R}^1 +S_\text{R}^2.
\end{eqnarray}

With the nonextensive thermodynamics as discussed above, the dS black hole spacetime can be described thermodynamically with the total entropy of the form as shown in Eq.~\eqref{Renyi com}, where $S_\text{R}^1$, $S_\text{R}^2$, and $S_\text{R}^{12}$ will be replaced later by the R\'enyi entropies of the black hole event horizon $S_{\text{R}(b)}$, the cosmic horizon $S_{\text{R}(c)}$, and the entire dS black hole spacetime $S$, respectively. In this paper, we investigate the thermodynamical stability and phase transition of the Sch-dS black hole in both separate and effective thermodynamical system approaches, based on the assumption that all subsystems and the entire system of the Sch-dS black hole are generally nonextensive. The dependence of thermodynamical behaviors of these systems in question on the nonextensive parameter $\lambda$ will be explored. It is also worthwhile to compare the results between the combination of all separated systems and the entire system as a whole. This can address in some levels about how the difference between these two approaches, namely, the separated and effective systems, relates to the nonextensivity effect associated with the correlations between these two horizons. Moreover, we discuss the comparison of the results between two aforementioned versions of the effective system approaches, namely, based on the interpretation of $M$ as the internal energy and as the enthalpy. 

Our paper is organized as follows. In Sec.~\ref{sect: Separated thermo}, we investigate thermodynamical properties of two separated systems, the black hole horizon and the cosmological horizon, using the R\'enyi entropy with different values of $\lambda$. In Sec.~\ref{sect: eff sys}, the entire thermodynamical system consisting of two horizons with different temperatures is considered in such a way that it is a single effective system. The investigation on thermodynamical stability in this section will be done through discussing the dependence of the effective R\'enyi temperature, thermodynamical volume, and the heat capacity on the ratio of $r_b$ to $r_c$. The phase structure is also explored through considering the dependence of the Gibbs free energy on the effective temperature. Importantly, these will be done with interpreting the mass as both internal energy and enthalpy.  Finally, we conclude with remarks of the effect of nonextensivity and the validity of our approach in giving a clear physical implication in Sec.~\ref{sect: conclusion}. In addition, the thermodynamical quantities for the system at the cosmic horizon and those for the effective system are discussed in Appendixes \ref{app: free energy at rc} and \ref{app: eff quan deriv}, respectively.


\section{Separated thermodynamical systems}\label{sect: Separated thermo}
Thermodynamics of Sch-dS spacetime have been intensively investigated. One of the key important issues in Sch-dS spacetime is that the black hole solution generally provides two distinct horizons, namely, black hole horizon and cosmic horizon. Therefore, thermodynamics of this kind of black hole can be treated as two thermodynamical systems. As a result, the investigation can be classified into two approaches; the thermodynamical systems are considered separately, and the thermodynamics is considered as an effective system. In this section, we will consider the first approach by assuming that the black hole and cosmic horizons are far enough to treat them as independent thermodynamical systems \cite{Tannukij:2020}. 

For the Sch-dS spacetime, the metric can be written as 
\begin{eqnarray}
	\text{d}s^2=-f(r)\text{d}t^2 + f(r)^{-1}\text{d}r^2 + r^2\text{d}\Omega^2,\hspace{1cm}
	f(r)=1-\frac{2M}{r}-\frac{\Lambda}{3}r^2. \label{metric}
\end{eqnarray}
The parameters $M$ and $\Lambda$ are the Arnowitt-Deser-Misner mass of black hole and the cosmological constant, respectively. For the positive value of $\Lambda$, it corresponds to the Sch-dS solution, while the negative value corresponds to the Sch-AdS solution. Generally, there are two event horizons: the black hole horizon denoted by $r_b$ and the cosmic horizon denoted by $r_c$ ($r_b$ is always not less than $r_c$). In other words, the possible values of $r_b$ are in the range $0\leq r_b\leq r_c$, while those of $r_c$ are in the range $r_b\leq r_c<\infty$. Using the horizon equations $f(r_b)=0$ and $f(r_c)=0$, the mass and cosmological constant at the horizons can be, respectively, expressed in terms of two horizons as
\begin{eqnarray}
	M&=&\frac{r_b r_c (r_b+r_c)}{2 \left(r_b^2+r_b r_c+r_c^2\right)}
	,\label{M}\\
	\Lambda&=&\frac{3}{r_b^2+r_b r_c+r_c^2}
	.\label{Lamb}
\end{eqnarray}
It is seen that both the mass and cosmological constant are always positive for any values of $r_b$ and $r_c$.

As we mentioned in the previous section, the thermodynamics of the black holes may respect the nonextensive system and then one has to consider a more proper statistics to evaluate the thermodynamics properties of the black holes. In the present work, we use the R\'{e}nyi statistics to examine the thermodynamics properties of the Sch-dS black hole. According to the R\'{e}nyi statistics, the nonextensive nature can be explained through the R\'{e}nyi entropy which is additive by construction. Therefore, the nonextensive system is  compatible with the zeroth law of thermodynamics. In other words, the empirical temperature of the system can be well defined. The R\'{e}nyi entropy $S_\text{R}$ can be expressed in terms of the Bekenstein-Hawking (BH) entropy $S_\text{BH}$ as \cite{Biro2011}
\begin{eqnarray}
	S_\text{R}=\frac{1}{\lambda}\ln\,(1+\lambda S_\text{BH}), \label{Renyi-entropy}
\end{eqnarray}
where $\lambda$ is the nonextensive parameter which is valid in the range $-\infty<\lambda<1$ \cite{Renyi:1959}. Note that the BH entropy at $r_b$ and $r_c$ are $\pi r_b^2$ and $\pi r_c^2$, respectively. Moreover, the R\'{e}nyi entropy can be reduced to the BH one by taking a limit $\lambda\to0$ called the Gibbs-Boltzmann (GB) limit. It is noted that the parameter $\lambda$ is valid for its whole range (from $-\infty$ to $1$), but the entropy $S_\text{R}$ is well defined when $1+\lambda S_\text{BH}>0$. Under this requirement, one obtains a condition for $\lambda$ as follows:
\begin{eqnarray}
	\lambda>-\frac{1}{\pi r_{b,c}^2}.
\end{eqnarray}
It is obvious that the entropy $S_\text{R}$ is always well defined for positive values of $\lambda$ ($0<\lambda<1$). However, for negative values of $\lambda$, its magnitude has to be very small when the horizon is very large (e.g., $r_c=10, |\lambda|\lesssim0.003$ for the system at the cosmic horizon). Therefore, we will consider only the positive values of $\lambda$ in the present paper. Note that, $\lambda>0$ is the necessary condition to obtain the locally stable-unstable phase transitions of the system on the black hole horizon \cite{Tannukij:2020}.

Let us consider the thermal system defined for each horizon of the black hole. In order to investigate the thermodynamics with respect to the R\'{e}nyi statistics, one recalls the first law of black hole thermodynamics as investigated in Refs.~\cite{Urano:2009xn,Kastor:2009wy}:
\begin{eqnarray}
	\text{d}M=\frac{\kappa_{(b,c)}}{2\pi}\,\text{d}(\pi r_{b,c}^2 )- \frac{1}{6} r_{b,c}^3  \,\text{d}\Lambda.
\end{eqnarray}
By identifying the temperature as $T_{(b,c)}=|\kappa_{(b,c)}|/2\pi $, entropy as $S= \pi r_{b,c}^2$, pressure  as $P = -\Lambda/8\pi$, and volumes as $V_{b,c}= 4\pi r_{b,c}^3/3$, one obtains the similar form of the first law of thermodynamics $\text{d}M=T\,\text{d}S+ V  \,\text{d}P$. It is obvious that the parameter $M=M(S,P)$ now plays the role of the enthalpy, and the internal energy is then written as $U=M-PV$. In this work, we adopt the same form of the first law of thermodynamics, $\text{d}M=T\,\text{d}S+V\,\text{d}P$ but now the entropy is described by the R\'{e}nyi entropy defined in Eq. (\ref{Renyi-entropy}). As a result, the R\'{e}nyi temperatures of both systems can be defined as follows:
\begin{eqnarray}
	T_{\text{R}(b)}=\left(\frac{\partial M}{\partial S_{\text{R}(b)}}\right)_\Lambda, \quad T_{\text{R}(c)}=-\left(\frac{\partial M}{\partial S_{\text{R}(c)}}\right)_\Lambda. \label{tem-R}
\end{eqnarray}
It is important to note that the expression of $T_{\text{R}(c)}$ with the minus sign is equivalent to the formulation derived from the surface gravity. This leads to the first law of thermodynamics as $\text{d}M=-T_{\text{R}(c)}\,\text{d}S_{\text{R}(c)}+ V_c \,\text{d}P$. The mass parameter $M$ is also treated as a function of the horizons, $M= M(r_b, r_c)$. Hence, the temperature defined in Eq. (\ref{tem-R}) can be found by fixing $\Lambda$ or, in other words, fixing the ratio of $r_b $ and $r_c$ in such a way that $\Lambda$ is kept constant. As a result, the temperatures are computed as
\begin{eqnarray}
	T_{\text{R}(b)}
	&=&\frac{(r_c-r_b)(2r_b+r_c)(1+\lambda\pi r_b^2)}{4\pi r_b(r_b^2+r_b r_c+r_c^2)},\label{TRb}\\
	T_{\text{R}(c)}
	&=&\frac{(r_c-r_b)(r_b+2r_c)(1+\lambda\pi r_c^2)}{4\pi r_c(r_b^2+r_b r_c+r_c^2)}.\label{TRc}
\end{eqnarray}
It is seen that $T_{\text{R}(b)}$ and  $T_{\text{R}(c)}$ is always positive because of the reality condition for the entropy which satisfies $1+\lambda\pi r_{r,c}^2>0$. The results in Eqs.~\eqref{TRb} and \eqref{TRc} are also obtained from the formula $T_{\text{R}(b,c)}=|\kappa_{(b,c)}|/2\pi=\frac{1}{4\pi}|\partial_rf|_{r=r_{b,c}}$. Furthermore, the temperature at the black hole horizon agrees with that in the literature [e.g., it can be checked by substituting $M$ and $\Lambda$ in Eqs.\eqref{M} and \eqref{Lamb} to the expression in Ref.~\cite{Tannukij:2020}].

\begin{figure}[h!]
\begin{center}
\includegraphics[scale=0.522]{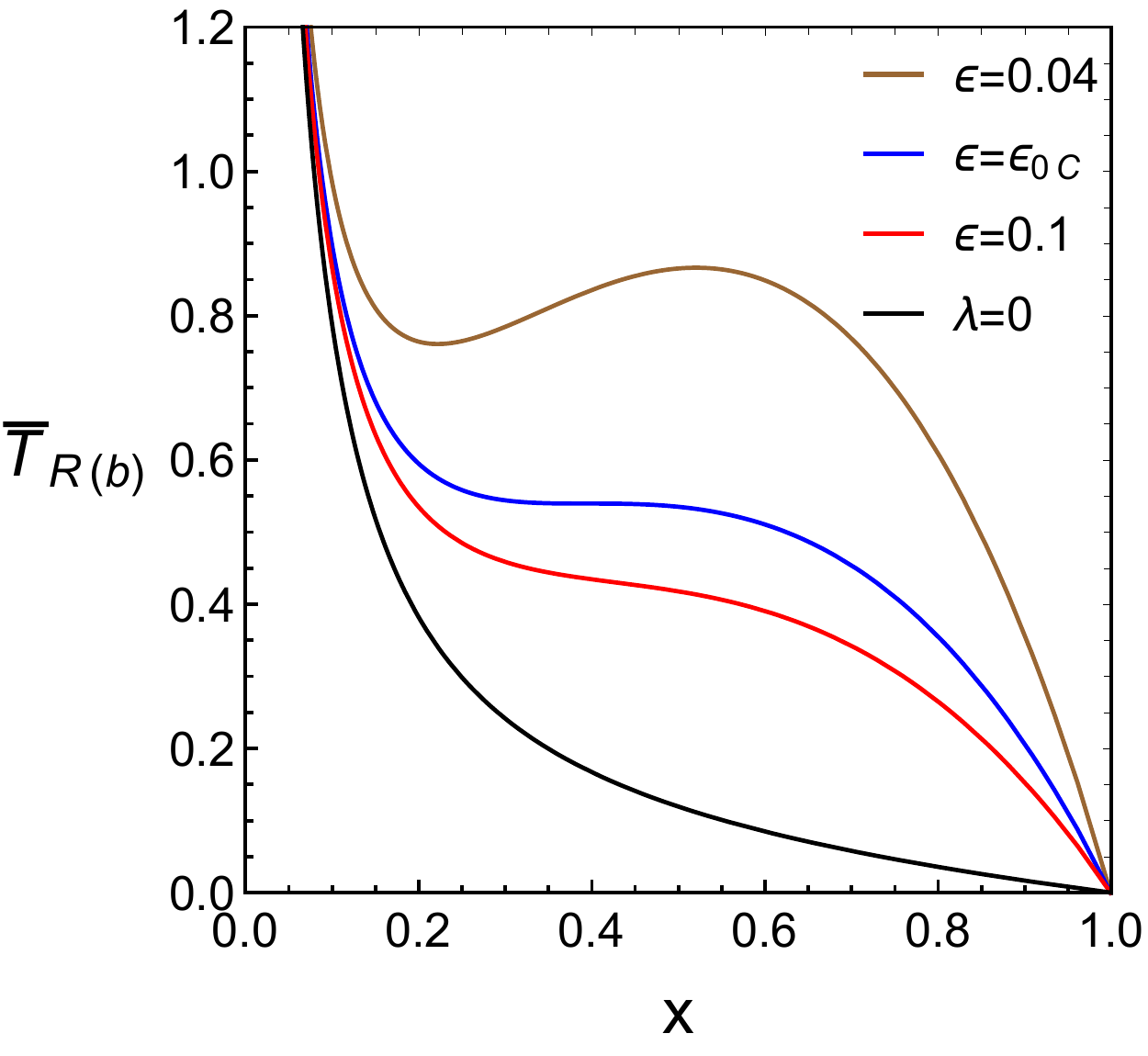}\hspace{1.cm}
\includegraphics[scale=0.5]{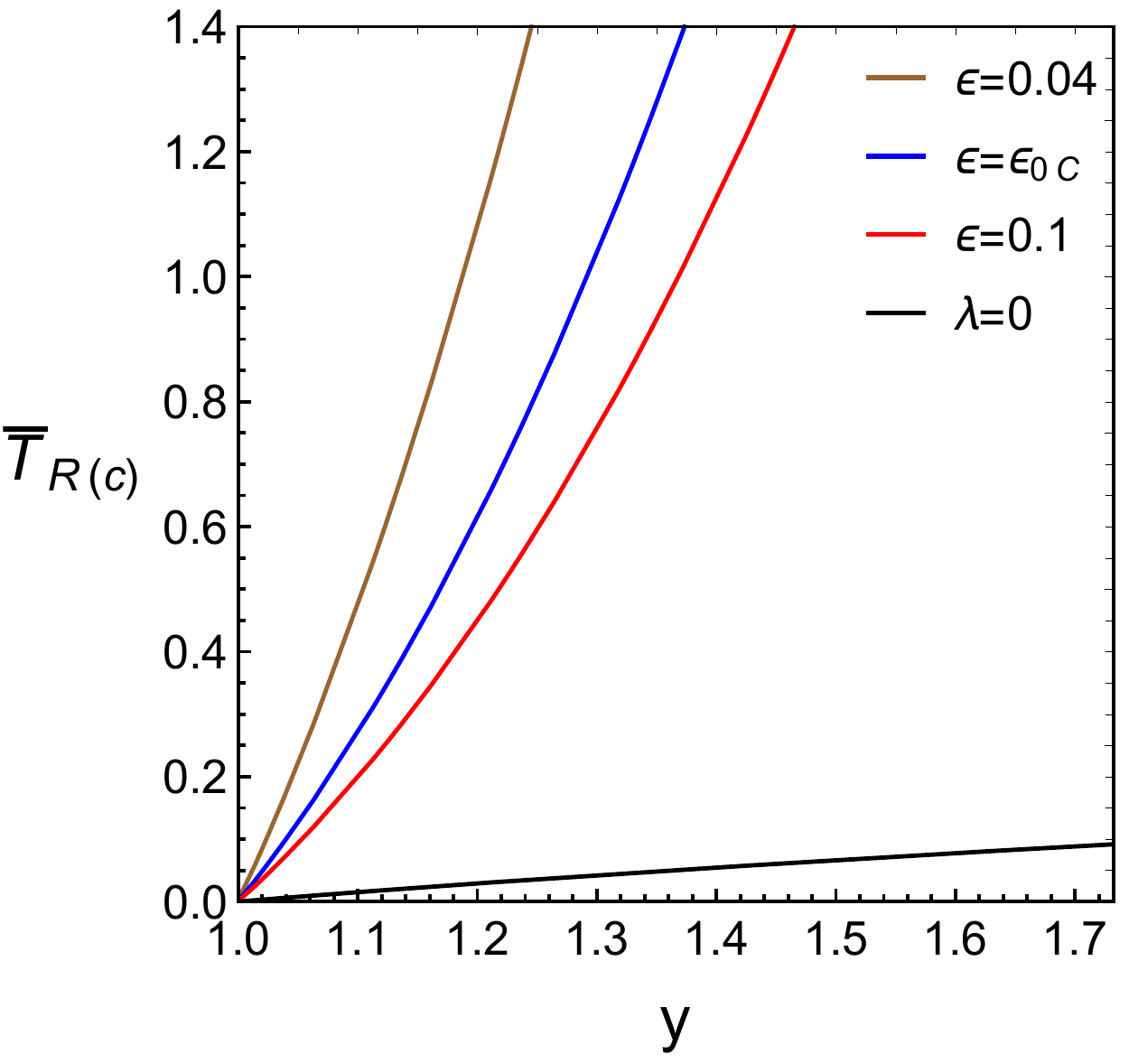}
\end{center}
\caption{The R\'{e}nyi temperatures $\bar{T}_{\text{R}(b,c)}=T_{\text{R}(b,c)}\big/\sqrt{\Lambda}$ for the systems at\\ the black hole horizon (left) and cosmic horizon (right) for various values of $\epsilon$\\ including the ones for the GB limit represented as black lines.}\label{fig:TRbTRc-L}
\end{figure}

For the black hole with multiple horizons, the systems can be described by only two horizon radii, which are $r_b$ and $r_c$. In the literature, the thermodynamical quantities are analyzed in terms of the ratio $r_b/r_c$ with fixing $r_c$. Unfortunately, this variable is not suitable for our case, because the feature of the temperature at a given $r_b/r_c$ does not imply the sign of the heat capacity at that point. As discussed in Ref.~\cite{Tannukij:2020}, the existence of the local minimum of the temperature profile infers the existence of the locally stable-unstable phase transition. This behavior can be obtained from analyzing the slope of the temperature, which is proportional to the heat capacity. The positivity of the heat capacity can be inferred from the positive slope of the temperature, and the heat capacity changes sign at the local extrema of the temperature. However, in our case, the temperature profile does not directly relate to the behavior of the heat capacity and then does not directly infer the phase transition. It is because the temperature depends on both $r_b$ and $r_c$, which are not fixed when evaluating the heat capacity. Hence, the change of the temperature in our case should be analyzed with fixing $P$ (or $\Lambda$). In order to analyze the behavior of the heat capacity from the temperature profile, one can write the temperature in terms of $\Lambda$. As a result, the temperatures for the system at the black hole and cosmic horizons can be, respectively, written as
\begin{eqnarray}
	T_{\text{R}(b)}
	&=&\frac{(1+\pi r_b^2\lambda)(1-\Lambda r_b^2)}{4\pi r_b} 
	=\sqrt{\Lambda}\frac{(x^2+\epsilon)(1-x^2)}{4\pi x\epsilon},\label{TRbL}\\
	T_{\text{R}(c)}
	&=&\frac{(1+\pi r_c^2 \lambda)(\Lambda r_c^2-1)}{4\pi r_c}
	=\sqrt{\Lambda}\frac{(y^2+\epsilon)(y^2-1)}{4\pi y\epsilon},\label{TRcL}
\end{eqnarray} 
where we have defined new set of dimensionless variables as $x\equiv r_b\sqrt{\Lambda}$, $y\equiv r_c\sqrt{\Lambda}$ and $\epsilon\equiv\Lambda/(\pi\lambda)$. From this form of the temperature profiles, it is obvious that both temperatures are positive where $1<r_c^2\Lambda<3$ and $0<r_b<r_c$ or $0<x<1$ and $1<y<\sqrt{3}$. The equation to solve for the extrema of the temperature $T_{\text{R}(b)}$ where $\Lambda$ is held fixed can be written in terms of dimensionless variables as 
\begin{eqnarray}
	3x^4-(1-\epsilon)x^2+\epsilon=0.\label{dTRL}
\end{eqnarray} 
For $\epsilon<1$, this equation is taken in the form of convex parabola with an argument of $x^2$, and its minimum point is at $x^2=(1-\epsilon)/6$. Requiring that the extrema must be two positive values leading to the minimum convex parabola must be negative, one can find the condition on $\epsilon$ as \cite{Tannukij:2020} 
\begin{eqnarray}
	\epsilon<\epsilon_{0C}=7-4\sqrt{3}\sim0.0718,\label{conT}
\end{eqnarray} 
which corresponds to $\lambda>\lambda_{0C}=(7+4\sqrt{3})\Lambda/\pi$. As a result, the extrema can be found by 
\begin{eqnarray}
	x^2_{\pm}=\frac{1}{6} \left(1-\epsilon  \pm \sqrt{\epsilon ^2-14 \epsilon +1}\right).\label{sol-dTRL}
\end{eqnarray}
Note that the equation to solve the extrema in Eq. (\ref{dTRL}) can be applied to the case of the cosmic horizon with replacing $x$ with $y$. Therefore, the minimum point is at $y^2=(1-\epsilon)/6$, which is out of the range $1<y<\sqrt{3} $. As a result, there is not an extremum point of the temperature for the system at the cosmic horizon. These behaviors can be seen explicitly from Fig. \ref{fig:TRbTRc-L}.

\begin{figure}[h!]
\begin{center}
\includegraphics[scale=0.5]{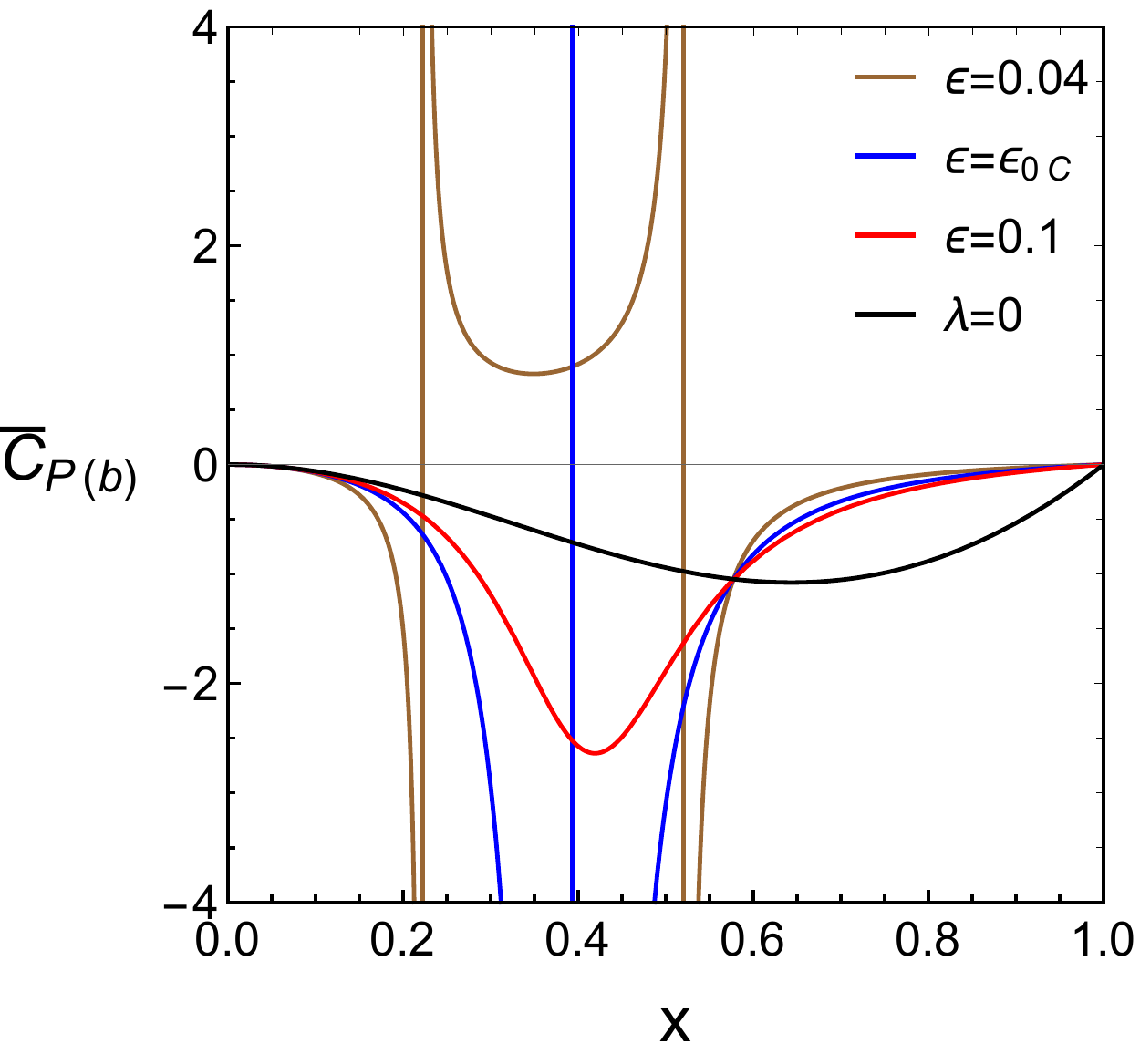}\hspace{0.5cm}
\includegraphics[scale=0.515]{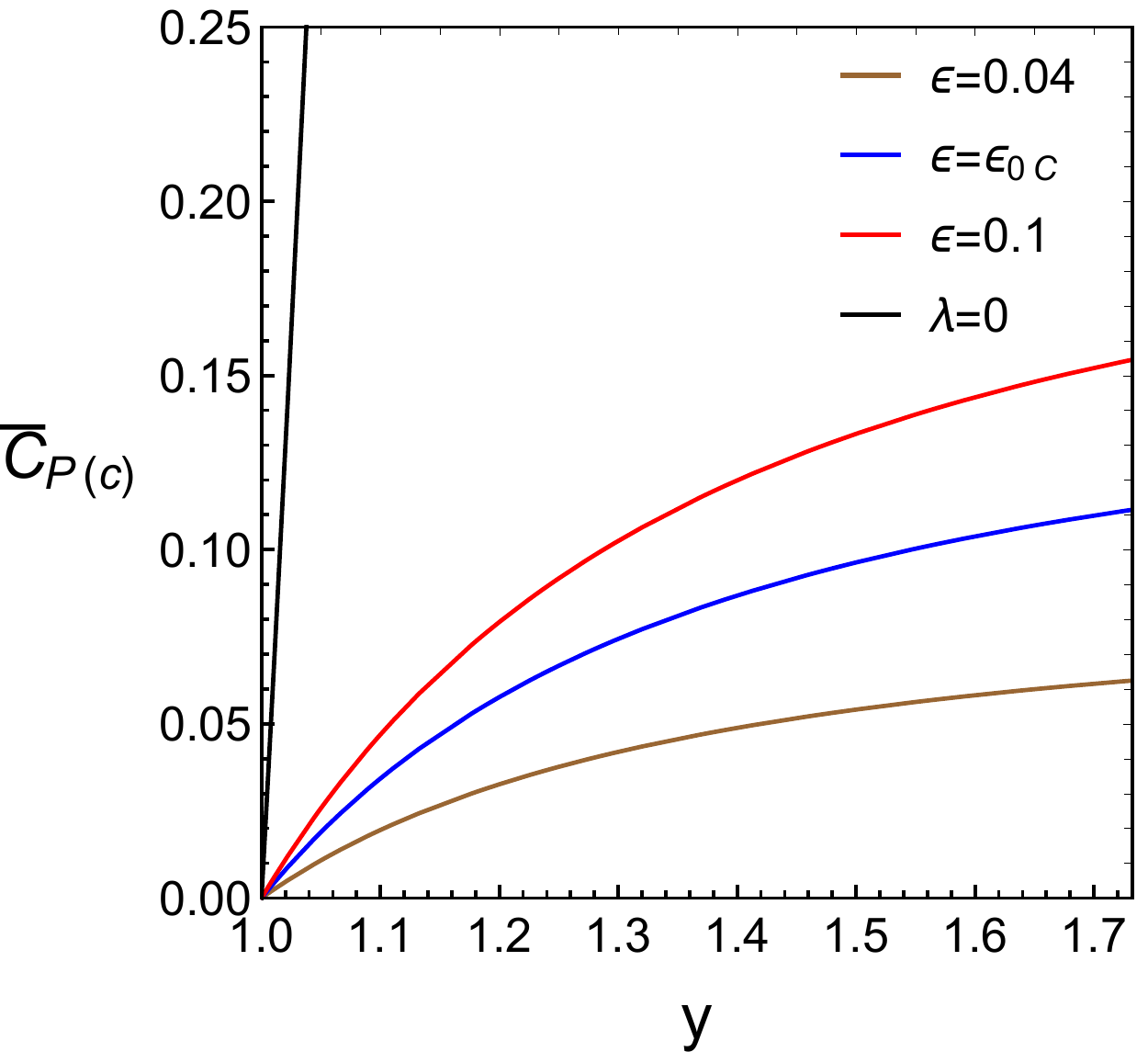}
\end{center}
\caption{The R\'{e}nyi heat capacity $\bar{C}_{P(b,c)}=\Lambda C_{P(b,c)}$ for the systems at the black hole horizon (left) and cosmic horizon (right) for various values of $\epsilon$ including the ones for the GB limit represented as black lines.}\label{fig:CPbCPc-L}
\end{figure}

Now let us move further to analyze the local stability of the black hole. The system with local stability is required to have a positive heat capacity. The thermal system with negative heat capacity will radiate thermal energy, and then the system gets hotter. In other words, the hotter the black hole is, the more it radiates, and then the system will vanish eventually. Therefore, the system with negative heat capacity is unstable in the sense that it can be formed but cannot live for long. With $M$ playing the role of the enthalpy, the heat capacity can be evaluated by fixing the pressure or $\Lambda$:
\begin{eqnarray}
	C_{P(b)}
	&=&\bigg(\frac{\partial M}{\partial T_{\text{R}(b)}}\bigg)_P = \frac{2\pi r_b^2 (\Lambda r_b^2 -1)}{3\pi \Lambda \lambda r_{b}^4 - (\pi \lambda - \Lambda)r_{b}^2 +1} 
	=\frac{2\pi x^2\left(x^2-1\right)\epsilon}{\Lambda\big[3x^4-(1-\epsilon)x^2+\epsilon\big]},\label{CPb}  \\
	C_{P(c)}
	&=&-\bigg(\frac{\partial M}{\partial T_{\text{R}(c)}}\bigg)_P = \frac{2\pi r_c^2 (\Lambda r_c^2 -1)}{3\pi \Lambda \lambda r_{c}^4 - (\pi \lambda - \Lambda)r_{c}^2 +1} 
	=\frac{2\pi y^2(y^2-1)\epsilon}{\Lambda\big[3y^4-(1-\epsilon)y^2+\epsilon\big]}.\label{CPc}
\end{eqnarray}
As we mentioned, the equation to have extrema in Eq. (\ref{dTRL}) is exactly the denominator in the expression for heat capacities in Eqs. (\ref{CPb}) and (\ref{CPc}). Therefore, the heat capacities diverge and change the signs at the extrema of the temperatures. For the $C_{P(c)}$, the numerator and denominator are always positive for the whole range of $1<y<\sqrt{3}$. For the $C_{P(b)}$, the numerator is always negative, and the denominator is a convex parabola with its minimum value being negative. Therefore, there are three ranges of $x$ which are positive for the middle one and negative for the others. The slope of the temperature is positive for $x$ which lies in between the two extrema (see the left panel in Fig. \ref{fig:TRbTRc-L}). The explicit behavior of the heat capacity can be illustrated in Fig. \ref{fig:CPbCPc-L}. Note that it is not possible to obtain a positive value of the heat capacity at the black hole horizon for the GB limit or $\lambda = 0$. This is also shown as the black line in the left panel in Fig. \ref{fig:CPbCPc-L}.

\begin{figure}[h!]
\begin{center}
\includegraphics[scale=0.32]{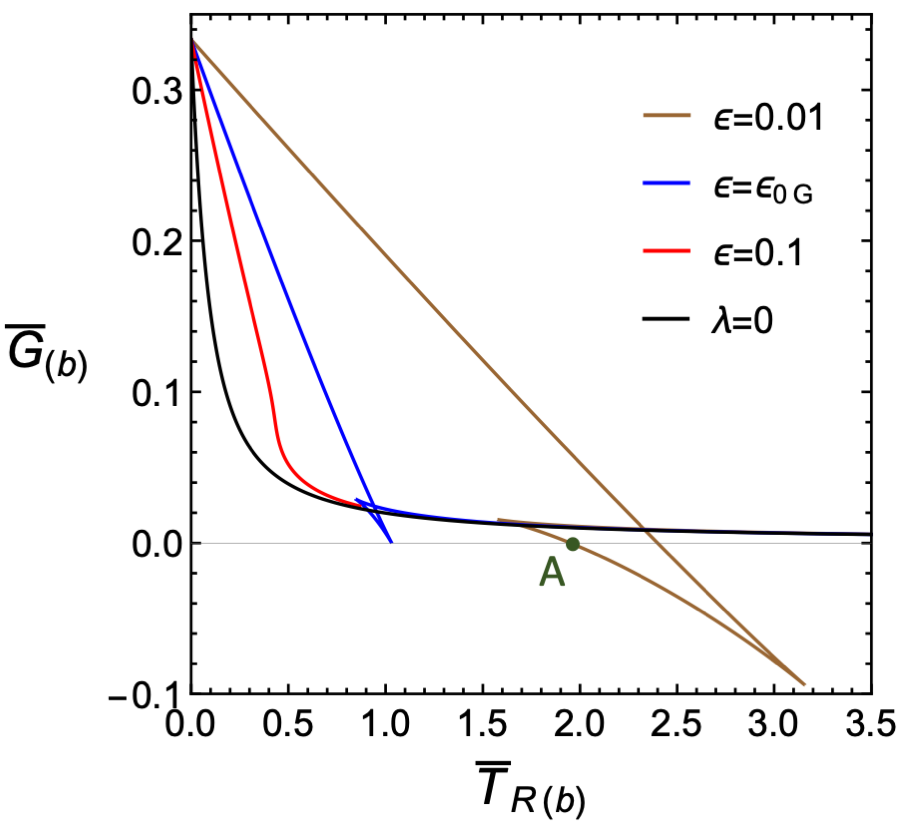}\hspace{0.5cm}
\includegraphics[scale=0.5]{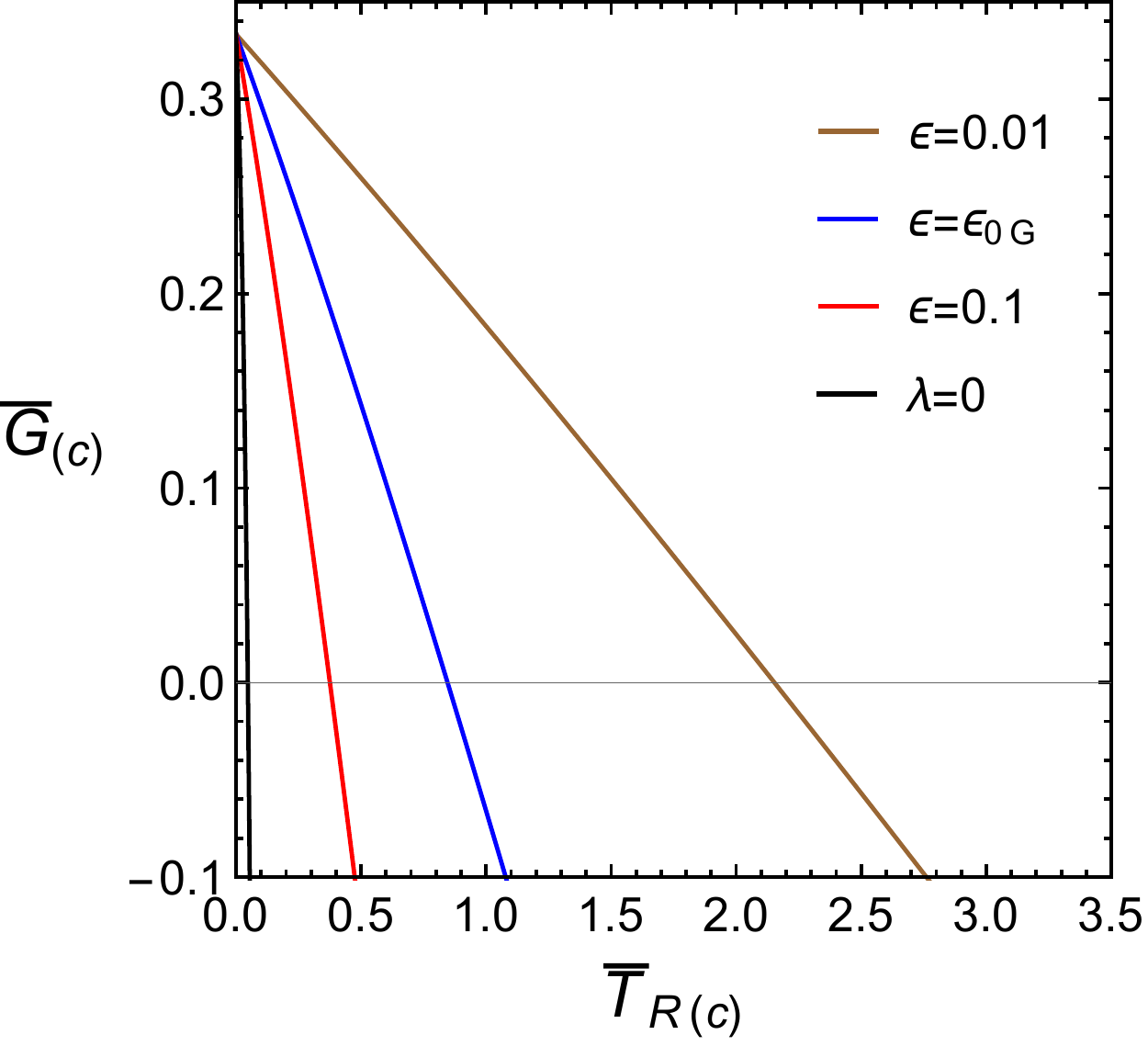}
\end{center}
\caption{The Gibbs free energy for the systems at the black hole horizon (left) and cosmic horizon (right) versus their own temperatures for various values of $\epsilon$ including the ones for the GB limit represented as black lines.}\label{fig:GbGc-L}
\end{figure}

We have already analyzed the local stability of the black hole by considering the behavior of the heat capacity. Now we will clarify the global stability by considering the Gibbs free energy. The thermodynamically stable system tends to prefer the system with lower free energy. In this sense, the black hole can be formed if the free energy of the system with the black hole is lower than that of the system without the black hole. In particular, since the free energy of the spacetime itself without the black hole is zero, the black hole can be formed if its Gibbs free energy is negative. The Gibbs free energy can be defined via the enthalpy $M$ as
\begin{eqnarray}
	G_{(b,c)}
	&=& M - T_{\text{R}(b,c)} S_{\text{R}(b,c)}\label{Gbc},\\ 
	\bar{G}_{(b)}=\sqrt{\Lambda} G_{(b)}
	&=&\frac{1}{12 x}\left[ 2 x^2 \left(3-x^2\right)-3 \left(1-x^2\right) \left(x^2+\epsilon \right) \ln \left(\frac{ x^2+\epsilon} {\epsilon }\right)\right],\label{Gb}\\
	\bar{G}_{(c)}=\sqrt{\Lambda} G_{(c)}
	&=&\frac{1}{12 y}\left[ 2 y^2 \left(3-y^2\right)-3 \left(y^2-1\right) \left(y^2+\epsilon \right) \ln \left(\frac{ y^2+\epsilon} {\epsilon }\right)\right].\label{Gc}
\end{eqnarray} 
As discussed in Ref.~\cite{Tannukij:2020}, the behavior of the free energy may be analyzed by the relation between $G$ and $T$ as $\big(\frac{\partial G}{\partial T}\big)_P=-S$. In our case, the slope $\Big(\frac{\partial G_{(b,c)}}{\partial T_{\text{R}(b,c)}}\Big)_P$ is always negative, since the entropy is always positive. Moreover, the function $G_{(b)}(T_{\text{R}(b)})$ at the black hole horizon is not smooth at the extrema of $T_{\text{R}(b)}$. These points are also marked as the locally stable-unstable phase transition, since they are the points where the heat capacity changes its sign. Actually, they correspond to the second-order phase transition, since the second derivative of the free energy which is proportional to the heat capacity, $\Big(\frac{\partial^2 G_{(b)}}{\partial T_{\text{R}(b)}^2}\Big)_P\propto C_{P(b)}$, diverges at these points. These features can be illustrated in Fig. \ref{fig:GbGc-L}. From the left panel in the figure, one can see that there are two cusps denoted by $x_\pm$ in Eq. (\ref{sol-dTRL}) for $\epsilon<\epsilon_{0C}$. The larger value of the solution ($x_{+}$) corresponds to the lower cusp. Therefore, one can evaluate a further upper limit of $\epsilon$ (or the lower limit of $\lambda$) by requiring that $G_{(b)}|_{x_+}<0$. It is not easy to find an analytical solution for this bound, since it contains a logarithmic function. One uses the numerical calculation to find this bound. As a result, the bound can be evaluated as 
\begin{eqnarray}
	\epsilon<\epsilon_{0G}\sim0.0328,\label{conG}
\end{eqnarray} 
which corresponds to $\lambda>\lambda_{0G}\sim\Lambda/(0.0328\pi)$. It can be seen explicitly from the left panel in Fig. \ref{fig:GbGc-L} that $G_{(b)}$ is always positive when $\epsilon>\epsilon_{0G}$. It is also inferred that the $G_{(b)}$ in the GB limit is always positive, since it is in the range $\epsilon>\epsilon_{0G}$. It is important to note that this bound is stronger than the one in Eq. (\ref{conT}), $\epsilon_{0G}<\epsilon_{0C}$ (or $\lambda_{0G}>\lambda_{0C}$). Hence, the system at the black hole horizon needs the condition of the nonextensive parameter $\lambda>\lambda_{0G}$ in order to be both locally and globally stable. Within this range of $\lambda$, it is possible to have the phase transition. Without the black hole, the free energy of the system is zero; this is supposed to be the free energy of the thermal radiation (or simply called hot gas) system. Now let us consider, for example, the left panel in Fig.~\ref{fig:GbGc-L} for $\epsilon = 0.01$; there exists a point denoted by ``A" such that the free energies of the hot gas and the black hole are the same. At this point, it is possible that the hot gas phase will change to a moderate-sized black hole phase. Since the slope of the free energy at the transition point is discontinuous, this is the first-order phase transition, the so-called Hawking-Page phase transition.

\begin{figure}[h!]
\begin{center}
\includegraphics[scale=0.55]{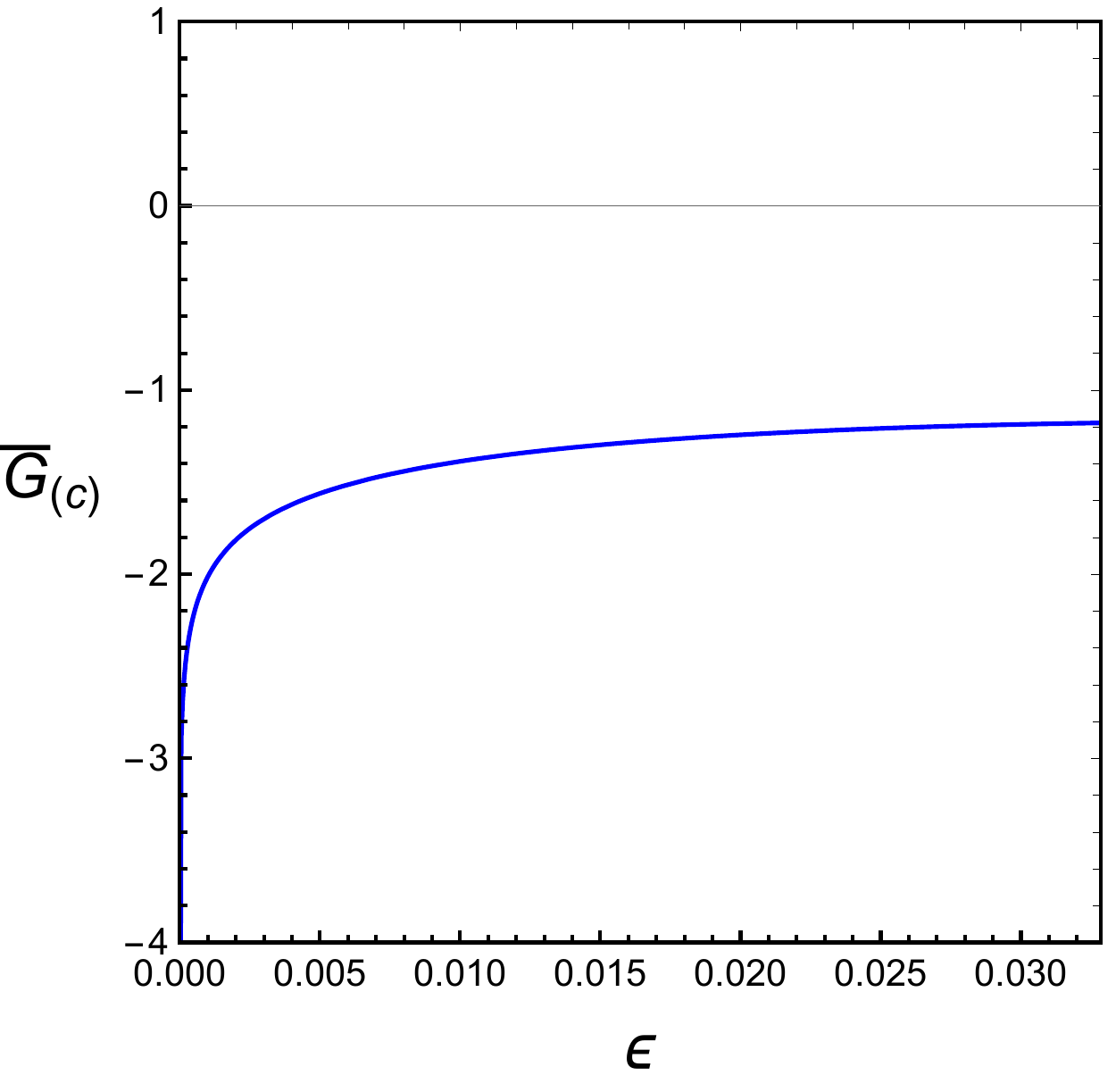}
\end{center}
\caption{The Gibbs free energy at cosmic horizon, $G_{(c)}(\epsilon)$ for $0<\epsilon<\epsilon_{0G}$. 
}\label{fig:Gc-rp}
\end{figure}

Since the Sch-dS black hole always has two horizons, it is required that both thermodynamical systems must be stable. Therefore, we have to check whether the free energy at the cosmic horizon is negative or not for the viable range from one evaluated at the black hole horizon. Since there is no extremum point for $T_{\text{R}(c)}$, there is no nonsmooth point in $G_{(c)}(T_{\text{R}(c)})$ as shown in the right panel in Fig.~\ref{fig:GbGc-L}. For a given value of $r_b$, it is possible to find $r_c$ in terms of $r_b$ and $\Lambda$ by using Eq. (\ref{Lamb}). As a result, one can find the value of $y = r_c\sqrt{\Lambda}$ corresponding to $x_{+}$ denoted by $y_{+}$ as follows:
\begin{eqnarray}
	y_+= \frac{1}{12} \left(3 \sqrt{2} \sqrt{23+\epsilon -\sqrt{(\epsilon -14) \epsilon +1}}-\sqrt{6} \sqrt{1-\epsilon +\sqrt{(\epsilon -14) \epsilon +1}}\,\right).\label{rcp}
\end{eqnarray}
One substitutes this value of $y$ to $\bar{G}_{(c)}$ in Eq. (\ref{Gc}) and then obtains $\bar{G}_{(c)} = \bar{G}_{(c)}(\epsilon)$. The expression is lengthy, so we do not put in here. Instead, it is more convenient to use a numerical plot of $\bar{G}_{(c)}$ as a function of $\epsilon$ as shown in Fig. \ref{fig:Gc-rp}. One can see that the locally and globally stable system at the black hole horizon has no problem, because $\bar{G}_{(c)}(\epsilon)$ is negative for the whole viable range of $\epsilon$, $ 0<\epsilon<\epsilon_{0G}$.

So far, we have investigated thermodynamical properties of each of the two event horizons, namely, the black hole horizon $r_b$ and the cosmic horizon $r_c$. The systems defined at $r_b$ and $r_c$ are characterized by different values of the R\'{e}nyi temperatures, $T_{\text{R}(b)}$ and $T_{\text{R}(c)}$, respectively. One may find that these black hole systems are inconclusive in terms of their thermal behaviors because of the different temperature profiles previously mentioned. One way to realize these systems as a single thermodynamical system is to consider them as one effective system, which will be carefully investigated in the next section.


\section{Effective system}\label{sect: eff sys}
In this section, the whole system is regarded as a single system called an effective system. The entropy of the effective system can be written as a total entropy as follows:
\begin{eqnarray}
	S=S_{\text{R}(b)}+S_{\text{R}(c)}
	=\frac{1}{\lambda}\Big[\ln\left(1+\lambda\pi r_b^2\right)+\ln\left(1+\lambda\pi r_c^2\right)\Big].\label{Seff Renyi}
\end{eqnarray}
Note that the negative values of $\lambda$ may give complex values of $S$. Since the entropy in Eq.~\eqref{Seff Renyi} is also expressed as
\begin{eqnarray}
	S=\frac{1}{\lambda}\ln\Big[\left(1+\lambda\pi r_b^2\right)\left(1+\lambda\pi r_c^2\right)\Big],
\end{eqnarray}
the condition for obtaining the well-defined entropy is
\begin{eqnarray}
	(1+\lambda\pi r_b^2)(1+\lambda\pi r_c^2)>0.
\end{eqnarray}
If $\lambda$ is negative, the above condition is satisfied when
	(i) $|\lambda|>\frac{1}{\pi r_b^2}$: $|\lambda|$ must be very large when $r_b$ is very small ($\lambda\to\infty$ as $r_b\to0$); and
	(ii) $|\lambda|<\frac{1}{\pi r_c^2}$: this is the same condition as the case of the separated system defined at the cosmic horizon. In order for the whole range of horizon radius to be allowed in the system, we will focus on the positive $\lambda$ in the present work. This also guarantees that the entropy is always positive. Moreover, as we have discussed in the previous section, it provides the possibility to have the phase transition for separated system consideration.

One of the crucial points for the effective system is that we can treat the parameter $M$ playing the role of either enthalpy or internal energy, while it is obscure to treat $M$ as the internal energy in the separated horizon approach. Therefore, we will divide our consideration into two parts depending on the thermodynamical role of the parameter $M$.

\subsection{$M$ as enthalpy}\label{MEn}
In this subsection, we consider the parameter $M = M(S,P)$ as the enthalpy of the system. This choice is quite natural, since it can be reduced to the separated horizon approach in the proper limit. Therefore, the results can be qualitatively compared to the separated horizon approach. In this sense, one may see how the system deviates from the separated horizon approach where the assumptions are relaxed, for example, in the case that the temperatures of two systems are not much different so that the heat transfer is supposed to be negligible.

We begin our consideration by expressing the differentiation of $M$ as the first law of thermodynamics as follows:
\begin{eqnarray}
	\text{d}M=T_\text{eff,(En)}\,\text{d}S+V_\text{eff}\,\text{d}P,\label{dM en}
\end{eqnarray}
where the pressure of this effective system is defined via $P=-\frac{\Lambda}{8\pi}$, which is analogous to that for separated systems in the previous section. From the fact that all state parameters $M$, $S$, and $P$ can be written in terms of only $r_b$ and $r_c$, the above relation leads to the expressions for the effective temperature of the system (the derivation is in Appendix~\ref{app: eff quan deriv}):
\begin{eqnarray}
	T_\text{eff,(En)}
	&=&\Big(\frac{\partial M}{\partial S}\Big)_P
	=\frac{\big(\frac{\partial M}{\partial r_b}\big)_{r_c}\big(\frac{\partial P}{\partial r_c}\big)_{r_b}-\big(\frac{\partial P}{\partial r_b}\big)_{r_c}\big(\frac{\partial M}{\partial r_c}\big)_{r_b}}{\big(\frac{\partial S}{\partial r_b}\big)_{r_c}\big(\frac{\partial P}{\partial r_c}\big)_{r_b}+\big(\frac{\partial P}{\partial r_b}\big)_{r_c}\big(\frac{\partial S}{\partial r_c}\big)_{r_b}}\nonumber\\
	&=&\frac{(r_c-r_b) (2 r_b+r_c) (r_b+2 r_c) \left(\pi  \lambda  r_b^2+1\right)\left(\pi  \lambda  r_c^2+1\right)}{4 \pi  \left(r_b^2+r_b r_c+r_c^2\right) \big[2 \pi  \lambda  r_b r_c \left(r_b^2+r_b r_c+r_c^2\right)+r_b^2+4 r_b r_c+r_c^2\big]}.\label{Teff en}
\end{eqnarray}
It is important to note that the definition of effective temperature in the present paper is different from ones in the literature. The crucial point is that we adopt the argument such that the change of entropy with respect to the cosmic horizon  is in the opposite way to the change with respect to the black hole horizon. Therefore, the change of entropy in our case can be expressed as $\text{d}S=\Big(\frac{\partial S}{\partial r_b}\Big)_{r_c}\text{d}r_b-\Big(\frac{\partial S}{\partial r_c}\Big)_{r_b}\text{d}r_c$ as seen in Eq. (\ref{dS}). Note that the definition of the effective quantities defined in this way allows us to avoid the singularity in effective temperature, while the effective temperature defined in the usual way is inevitable to diverge at some point of a nonextensive parameter. 

\begin{figure}[ht!]
\begin{center}
\includegraphics[scale=0.5]{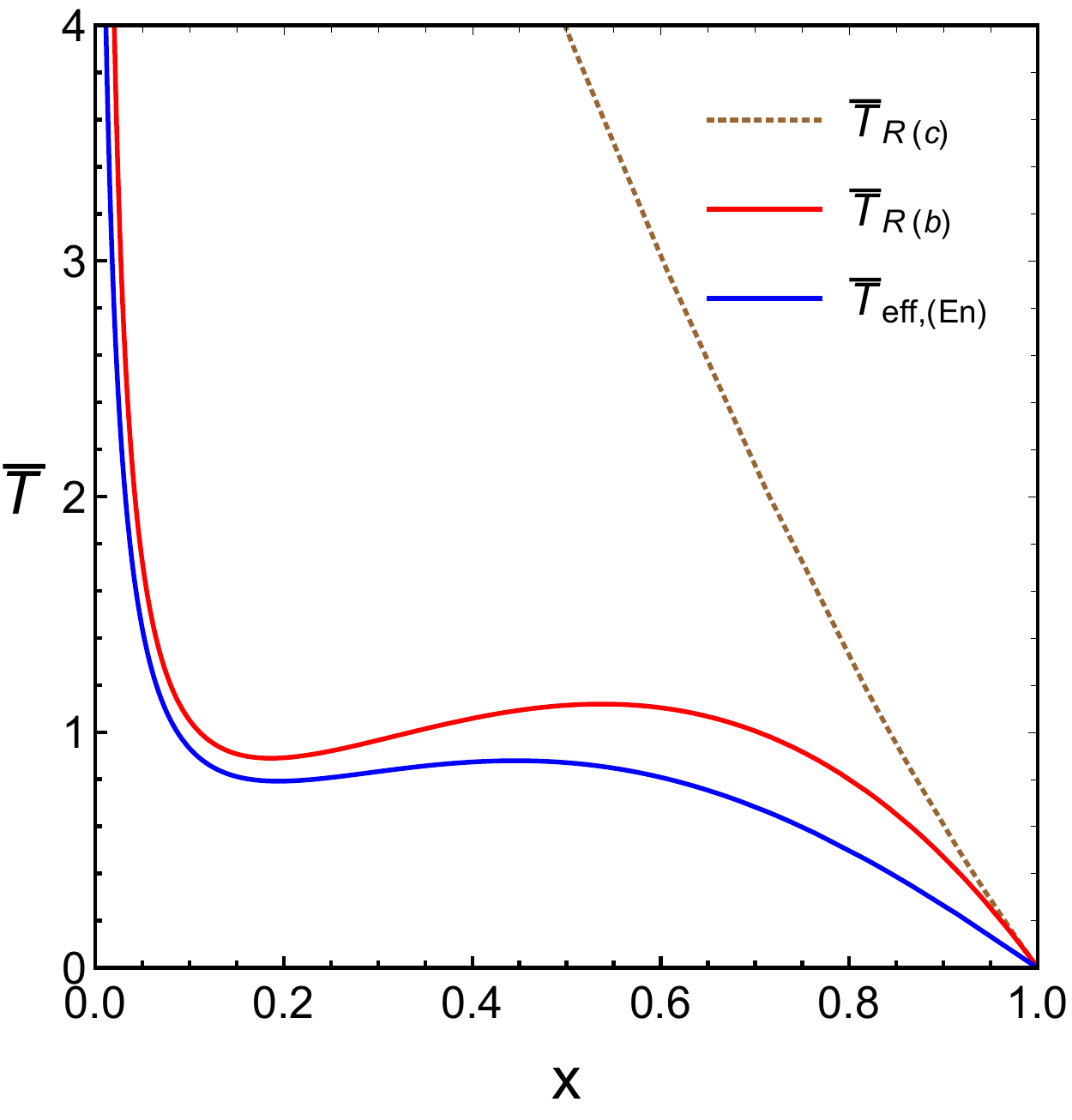}\hspace{1.cm}
\includegraphics[scale=0.353]{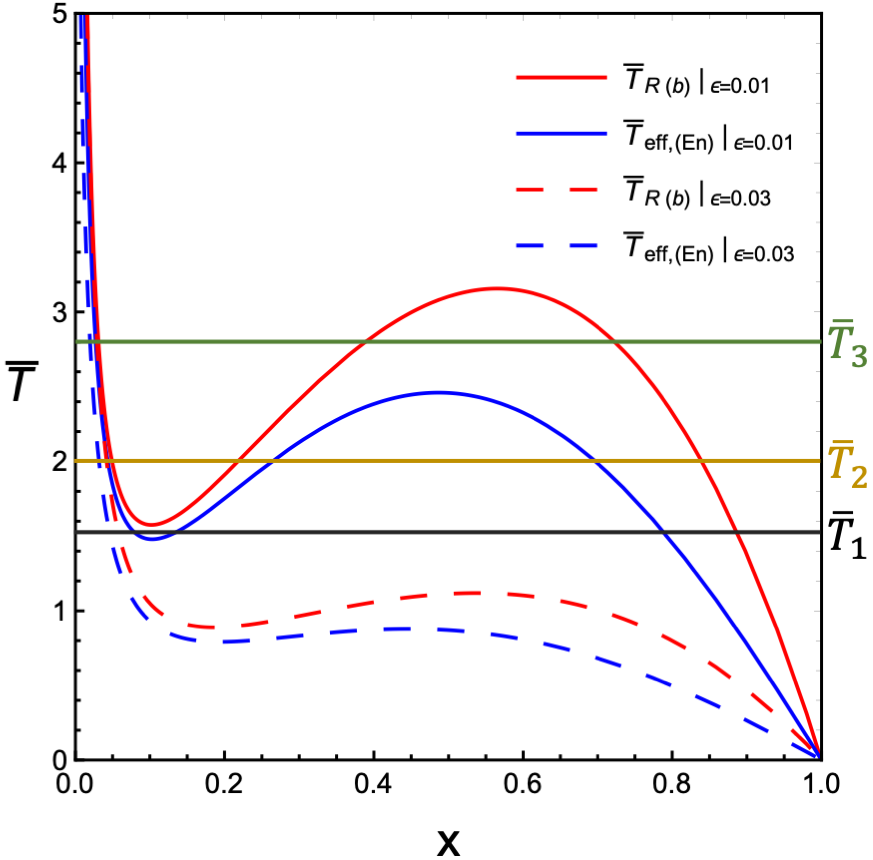}
\end{center}
\caption{The effective temperature $T_{\text{eff,(En)}}$ compared to \\the black hole temperature $T_{\text{R}(b)}$ and the temperature at cosmic horizon $T_{\text{R}(c)}$. \\The left panel includes those three temperatures with $\epsilon = 0.03$. \\The right panel includes $T_{\text{eff,(En)}}$ (blue) and $T_{\text{R}(b)}$ (red) with various values of $\epsilon$. }\label{fig:Teff-bc(En)}
\end{figure}

As we have analyzed in the previous section, the expression of the temperature in Eq. (\ref{Teff en}) is not suitable to analyze the behavior of the phase transition. This effective temperature can be rewritten in terms of $\Lambda$ as follows:
\begin{eqnarray}
	T_\text{eff,(En)}
	&=&\frac{\left(1-\Lambda  r_b^2\right)\left(\Lambda  r_c^2-1\right)\left(\pi  \lambda  r_b^2+1\right)\left(\pi  \lambda  r_c^2+1\right) }{4 \pi  \big[r_c \left(\pi  \lambda  r_b^2+1\right) \left(1-\Lambda  r_b^2\right)+r_b \left(\pi  \lambda  r_c^2+1\right) \left(\Lambda  r_c^2-1\right)\big]}.\label{Teff-en-L}
\end{eqnarray}
Note that, in principle, we can express $T_\text{eff,(En)}$ in such a way that $T_\text{eff,(En)}=T_\text{eff,(En)}(r_b,\Lambda)$ or $T_\text{eff,(En)}= T_\text{eff,(En)}(r_c,\Lambda)$ by using Eq. (\ref{Lamb}). We also found that the effective temperature is related to the temperatures at the black hole and cosmic horizons as follows
\begin{eqnarray}
	\frac{1}{T_\text{eff,(En)}}
	=\Big(\frac{\partial S}{\partial M}\Big)_P
	=\Big(\frac{\partial S_{\text{R}(b)}}{\partial M}\Big)_P-\Big(\frac{\partial S_{\text{R}(c)}}{\partial M}\Big)_P
	=\frac{1}{T_{\text{R}(b)} }+\frac{1}{T_{\text{R}(c)} }. \label{re-T}
\end{eqnarray}
Note that the effective temperature defined in this sense is not the temperature at the equilibrium state between two systems. Instead, this temperature is the representative quantity characterizing a single system called the effective system. Moreover, it is found that the effective temperature reduces to the black hole temperature for $r_c \rightarrow \infty$ and reduces to one at the cosmic horizon for $r_b \rightarrow 0$:
\begin{eqnarray}
	\lim_{r_c\to\infty}T_\text{eff,(En)}
	&=&T_{\text{R}(b)}=\frac{(1+\pi r_b^2\lambda)(1-\Lambda r_b^2)}{4\pi r_b},\label{TEn rc infty}\\
	\lim_{r_b\to 0}T_\text{eff,(En)}
	&=&T_{\text{R}(c)}=\frac{(1+\pi r_c^2 \lambda)(\Lambda r_c^2-1)}{4\pi r_c}.
\end{eqnarray}
This is a useful property since the effective quantity from two systems can be reduced to one of them in its proper limit. As a result, the behavior of the effective temperature profile is inherited from the black hole temperature such that there exist the local extrema as shown in Fig. \ref{fig:Teff-bc(En)}. From this figure, one can see that the profiles of $T_{\text{eff,(En)}}$ and $T_{\text{R}(b)}$ are tracked in similar locus for small $x$, since the contribution from $T_{\text{R}(c)}$ in Eq. (\ref{re-T}) is very small. Moreover, it is found that for fixing the radius of the black holes the effective temperature is always less than the temperature of the black holes determined from the separated horizon approach, $T_{\text{eff,(En)}}<T_{\text{R}(b)}$. As shown in the right panel in Fig. \ref{fig:Teff-bc(En)}, at temperature $\bar{T}_2$, the locally stable system in the effective description is always larger than that in the separated description. This implies that the thermodynamically stable black hole in the effective approach is always larger than the one in the separated horizon approach. Moreover, there exists, for example, the temperature $\bar{T}_1 (\bar{T}_3)$ in the right panel in Fig. \ref{fig:Teff-bc(En)} at which only a black hole in the effective (separated) system approach is stable. As a result, these particular temperatures can be used to distinguish between two approaches of black hole thermodynamics. For example, if we observe a black hole with temperature $\bar{T}_1 (\bar{T}_3)$, it can be argued that the effective (separated) system approach is the more reliable one.
\begin{figure}[ht!]
\begin{center}
\includegraphics[scale=0.5]{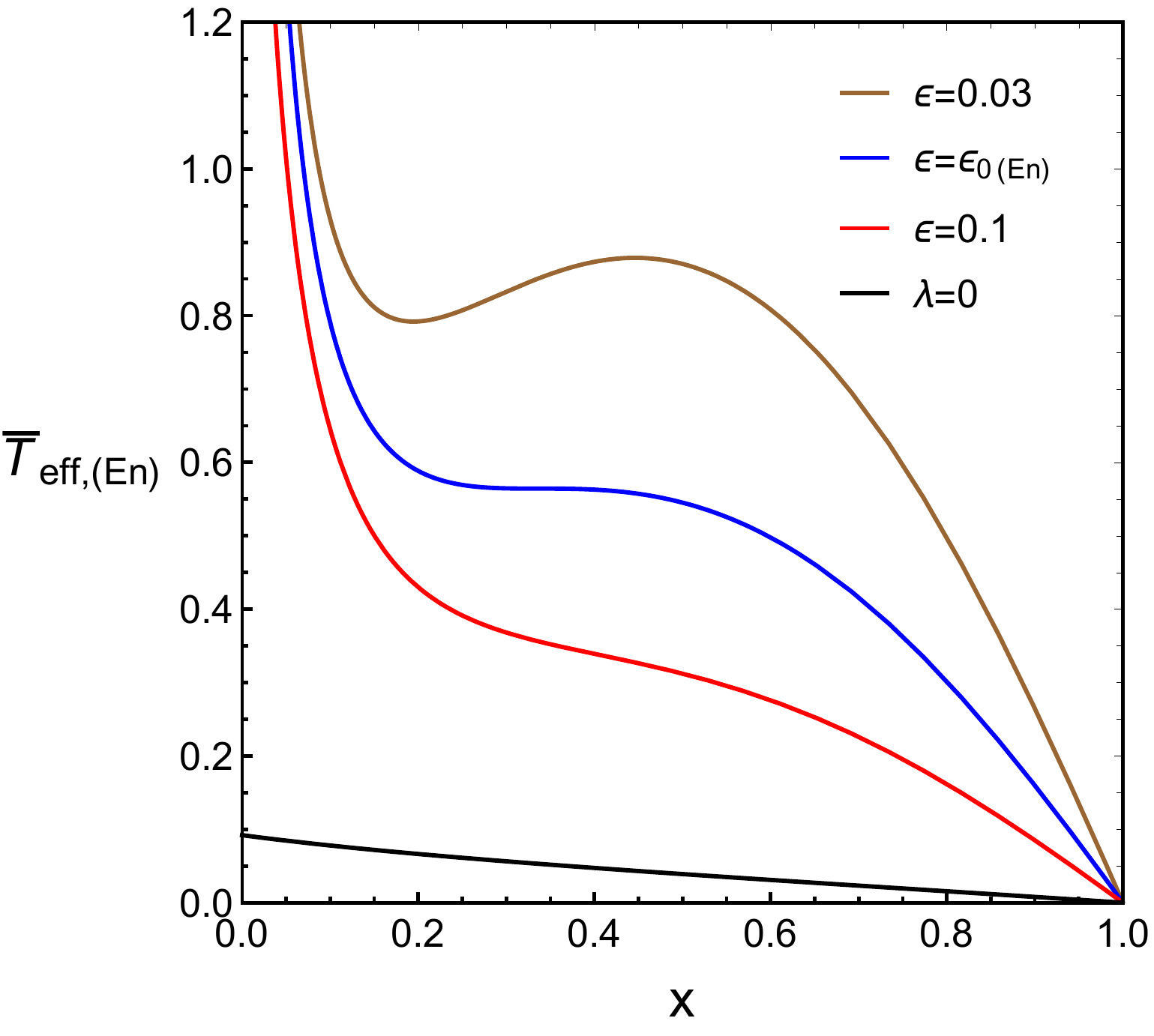}\hspace{.5cm}
\end{center}
\caption{The effective temperature $T_{\text{eff,(En)}}$ with various values of $\epsilon$ \\including the one for the GB limit represented as the black line.}\label{fig:Teff-ep(En)}
\end{figure}

By using the same way as we find the bound on the parameter $\epsilon$ in the separated horizon approach, one can find the upper limit of $\epsilon$ by finding solutions of $\text{d}T_{\text{eff,(En)}} =0$ in such a way that $\Lambda$ is held fixed. As a result, the equation can be written as
\begin{eqnarray}
	0&=&\left(3 x^4+x^2 (\epsilon -1)+\epsilon \right)+Y,\label{dTReff}\\
	Y&=&\frac{(1-x^2)^3 (x^2+\epsilon )^2 \left(3 y^4+y^2 (\epsilon -1)+\epsilon \right)}{\left(y^2-1\right)^3 \left(y^2+\epsilon \right)^2}.
\end{eqnarray}
From this equation, one can see that it is the equation for the separated horizon approach in Eq. (\ref{dTRL}) plus a positive small function $Y$. Therefore, the above equation is still a convex function. One can find the local minimum which depends only on $\epsilon$ and then find the value of the function at the minimum in terms of $\epsilon$. Ultimately, one can find the condition of $\epsilon$ by requiring that the value of the function in Eq.~\eqref{dTReff} at the minimum must be less than zero. By following the mentioned step, we found the upper bound of $\epsilon$ as 
\begin{eqnarray}
	\epsilon=\epsilon_\text{0(En)}=0.0507.
\end{eqnarray}
This value of the bound is lower than the one in the separated approach, since the contribution from $Y$ is positive. Note that we have to write $y$ in Eq. (\ref{dTReff}) in terms of $x$ through Eq. (\ref{Lamb}) and then solve the equation for $x$. This behavior can be seen explicitly in Fig. \ref{fig:Teff-ep(En)}. From this figure, one can see that there are no local extrema of $T_{\text{eff,(En)}}$ for $\epsilon>\epsilon_\text{0(En)}$.

\begin{figure}[ht!]
\begin{center}
\includegraphics[scale=0.5]{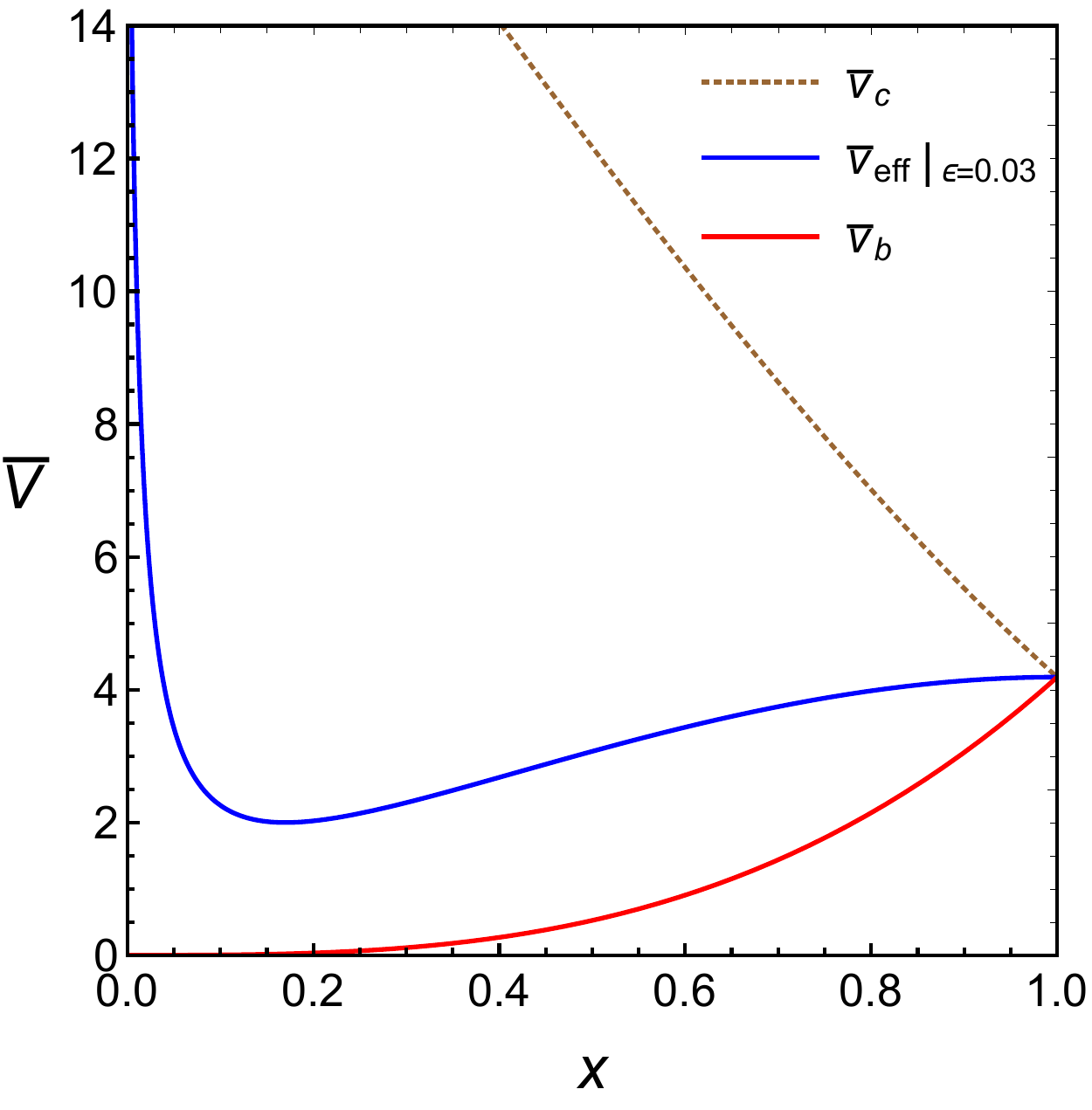}\hspace{.5cm}
\includegraphics[scale=0.525]{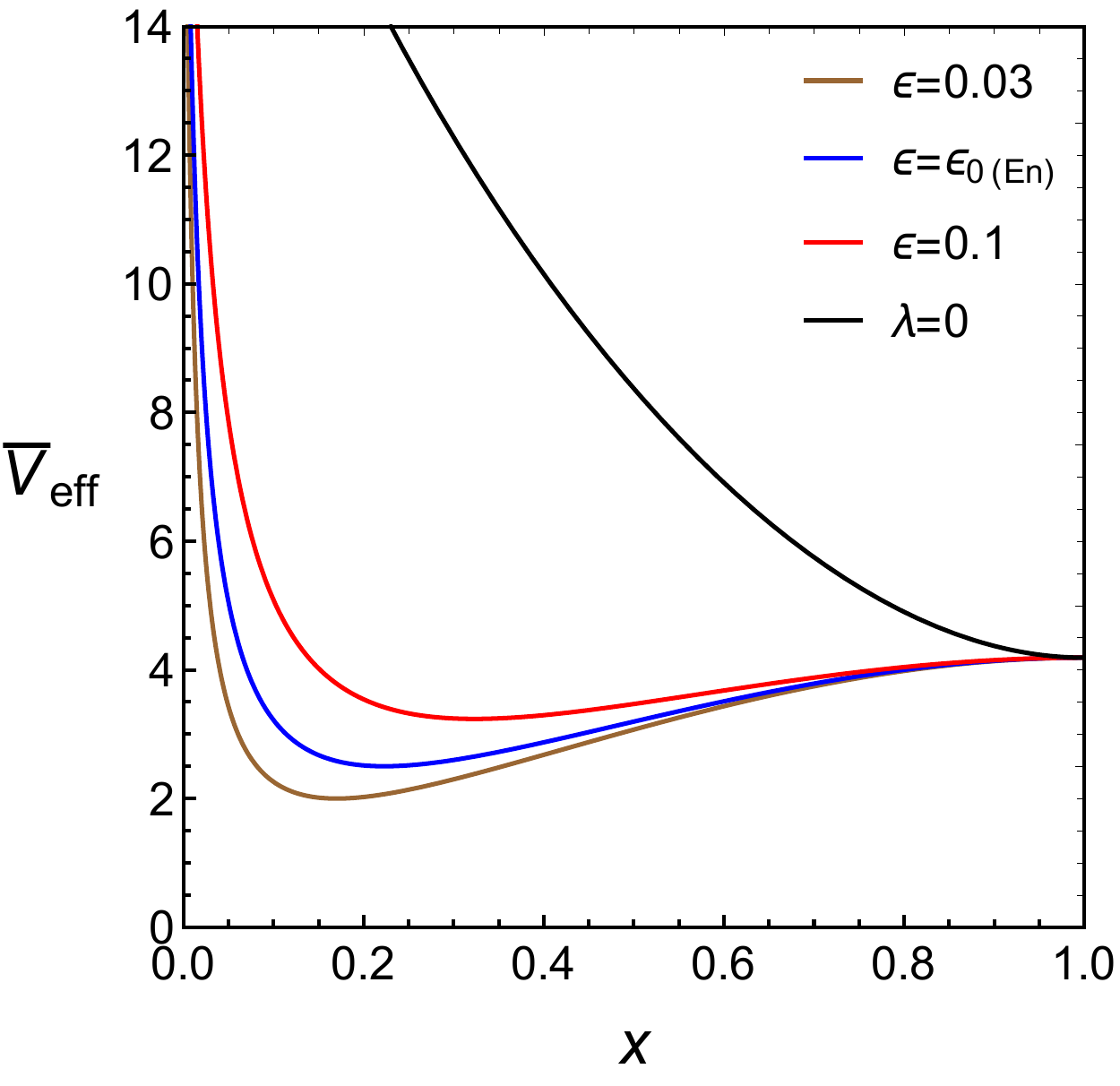}
\end{center}
\caption{The effective volume $V_{\text{eff}}$ compared to the black hole volume $V_{b}$ and the volume at cosmic horizon $V_{c}$ ($\bar{V}=\Lambda^{3/2}V$). The left panel includes those three volumes with $\epsilon = 0.03$. The right panel shows $V_{\text{eff}}$ with various values of $\epsilon$ including the one in the GB limit represented as the black line.}\label{fig:Veff-bc}
\end{figure}

The effective volume satisfying Eq.~\eqref{dM en} can be computed as
\begin{eqnarray}	
	V_\text{eff}
	&=&\Big(\frac{\partial M}{\partial P}\Big)_S
	=\frac{\big(\frac{\partial M}{\partial r_b}\big)_{r_c}\big(\frac{\partial S}{\partial r_c}\big)_{r_b}+\big(\frac{\partial S}{\partial r_b}\big)_{r_c}\big(\frac{\partial M}{\partial r_c}\big)_{r_b}}{\big(\frac{\partial S}{\partial r_b}\big)_{r_c}\big(\frac{\partial P}{\partial r_c}\big)_{r_b}+\big(\frac{\partial P}{\partial r_b}\big)_{r_c}\big(\frac{\partial S}{\partial r_c}\big)_{r_b}}\nonumber\\
	&=&\frac{4 \pi  \big[r_b^4 \left(r_b+2 r_c\right) \left(1+\pi  \lambda  r_c^2\right)+r_c^4 \left(r_c+2 r_b\right) \left(1+\pi  \lambda  r_b^2\right)\big]}{3 \big[ r_b \left(r_b+2 r_c\right) \left(1+\pi  \lambda  r_c^2\right)+r_c \left(r_c+2 r_b\right) \left(1+\pi  \lambda  r_b^2\right)\big]}\nonumber\\
	&=&T_\text{eff,(En)}\left(\frac{V_b}{T_{\text{R}(b)}}+\frac{V_c}{T_{\text{R}(c)}}\right).
\end{eqnarray}
From this equation, one can see that the effective volume can be interpreted as an average quantity weighted by $T_\text{eff,(En)}/T_{\text{R}(b,c)}$. Therefore, it provides a proper limit such that 
\begin{eqnarray}
	\lim_{r_c\to 0}V_\text{eff}&=&V_b=\frac{4\pi r_b^3 }{3},\\
	\lim_{r_b\to 0}V_\text{eff}&=&V_c=\frac{4\pi r_c^3 }{3}.
\end{eqnarray}
This is a nice property in the sense of thermodynamics quantities. According to the above results, the effective volume is a representation of the separated ones which are independent to each other. In other words, if one of the separated volumes vanishes, the effective volume will become another volume. 
Moreover, the effective volume for $r_b = r_c$ is automatically reduced to the thermodynamical volume of the extremal black hole. The behavior of the effective volume can be seen explicitly from Fig. \ref{fig:Veff-bc}.

\begin{figure}[ht!]
\begin{center}
\includegraphics[scale=0.5]{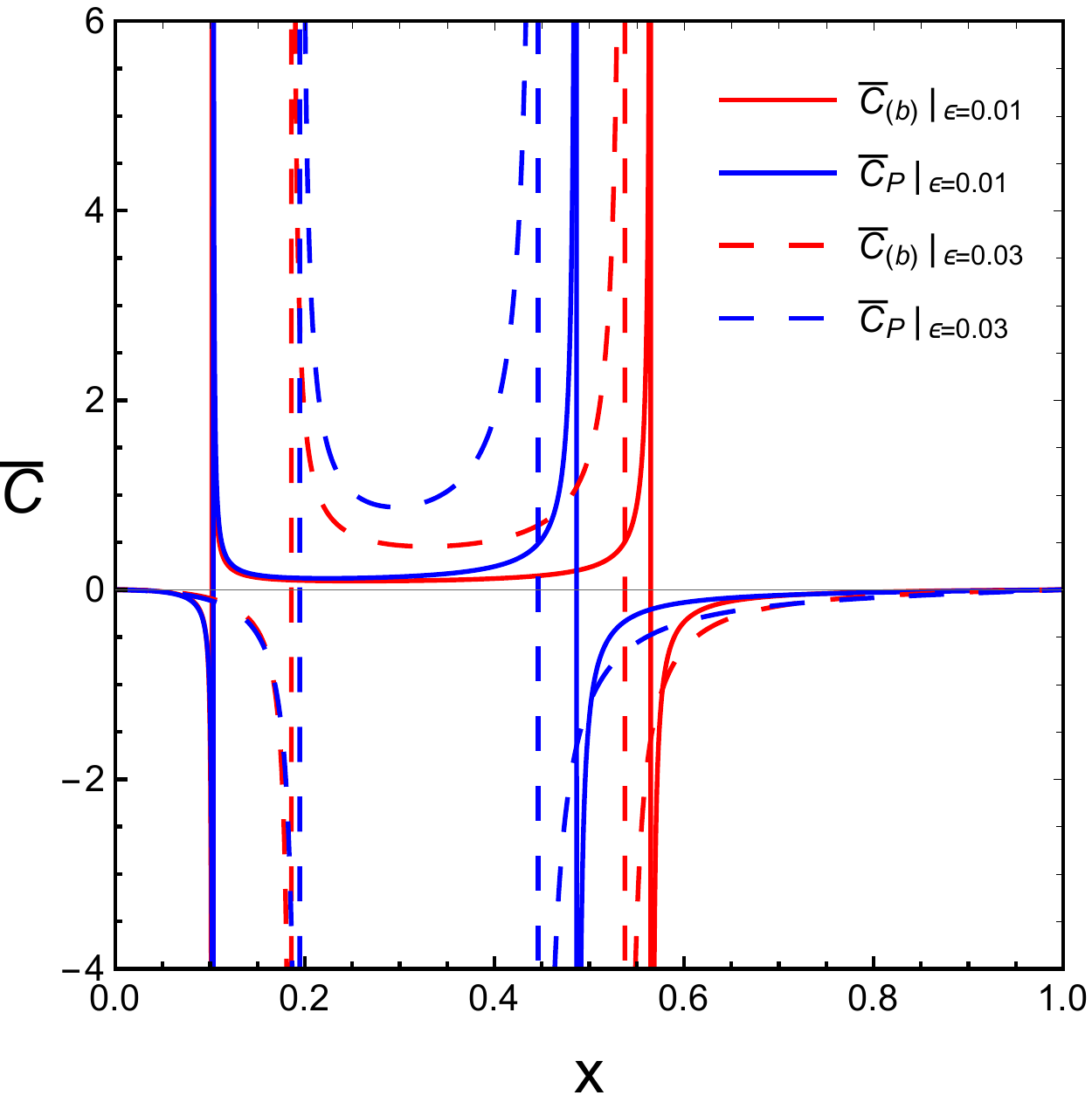}\hspace{.5cm}
\includegraphics[scale=0.517]{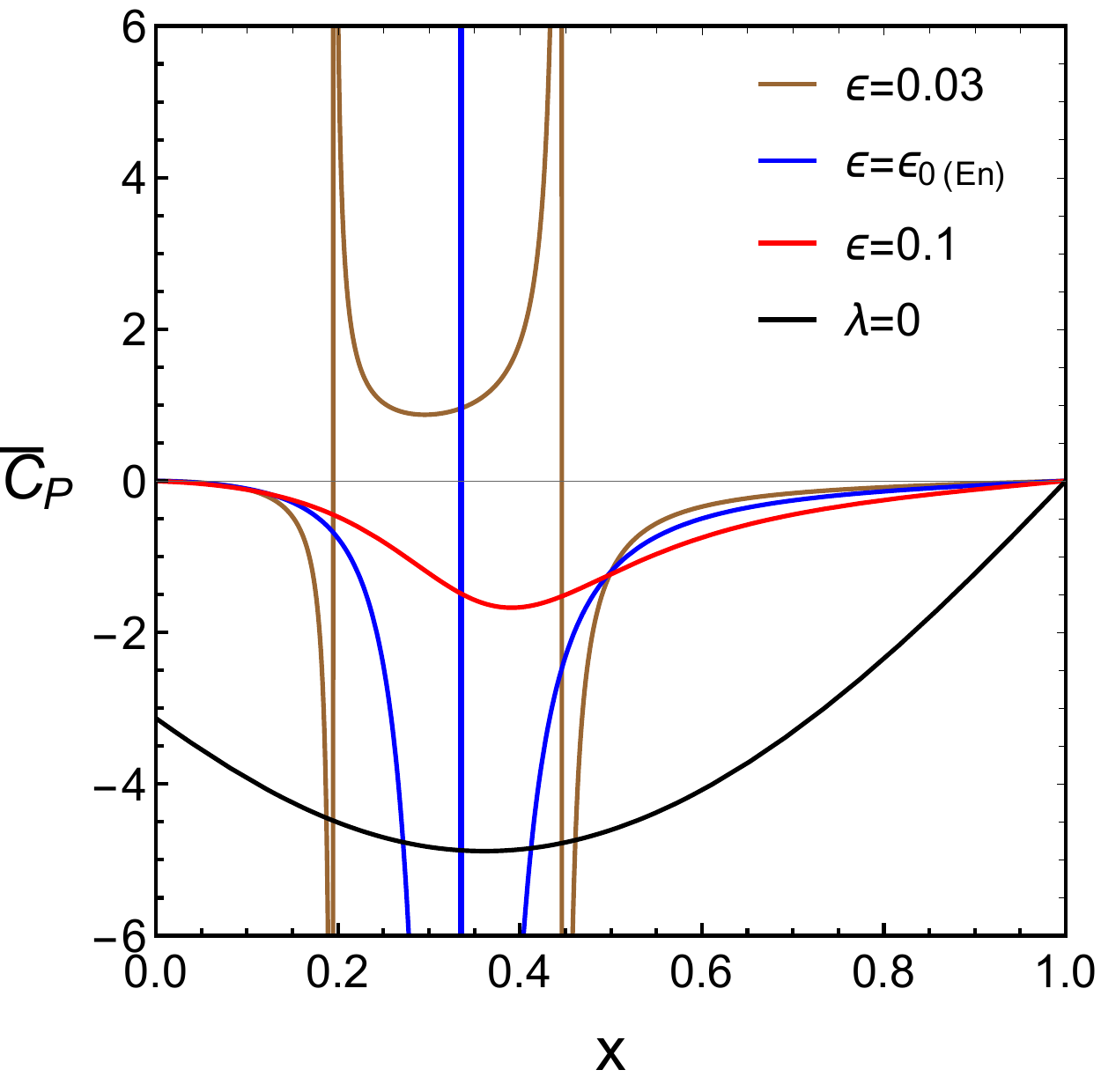}
\end{center}
\caption{The heat capacity at constant pressure in the effective system $C_P$ compared to the black hole heat capacity $C_{(b)}$. The left panel includes both heat capacities for the effective description (blue lines) and at the black hole horizon for the separated approach (red lines) with $\epsilon=0.01$ and $\epsilon=0.03$. The right panel includes $C_P$ for various values of $\epsilon$ including the one in the GB limit represented as the black line.}\label{fig:CP}
\end{figure}

The heat capacity at constant pressure is defined as 
\begin{eqnarray}
	C_P
	&=&\left(\frac{\partial M}{\partial T_\text{eff,(En)}}\right)_P,\nonumber\\
	\bar{C}_P= \Lambda C_P &=&-\frac{2 \pi  \epsilon \left(1-x^2\right)  (x-y)^2 (x y (\epsilon +2)+\epsilon )^2}{\left(y^2-1\right)^2 \left(y^2+\epsilon \right)^2 \big[\big\{3 x^4+x^2 (\epsilon -1)+\epsilon \big\}+Y\big]}.
\end{eqnarray}
Note that the above expression is written in terms of the dimensionless variables. From this equation, one can see that the heat capacity diverges at the extrema of the temperature, since there exists a convex function in the denominator which is the same as one in Eq. (\ref{dTReff}). As a result, according to the slope of the temperature, there are three parts of the heat capacity; the middle one is positive, and the others are negative as shown in Fig. \ref{fig:CP}. As a result, the moderate-sized black hole is thermodynamically locally stable within a range of parameter $\epsilon<\epsilon_\text{0(En)}\sim 0.0507$ [or $\lambda>\lambda_\text{0(En)}\sim\Lambda/(0.0507\pi)$]. 

\begin{figure}[ht!]
\begin{center}
\includegraphics[scale=0.5]{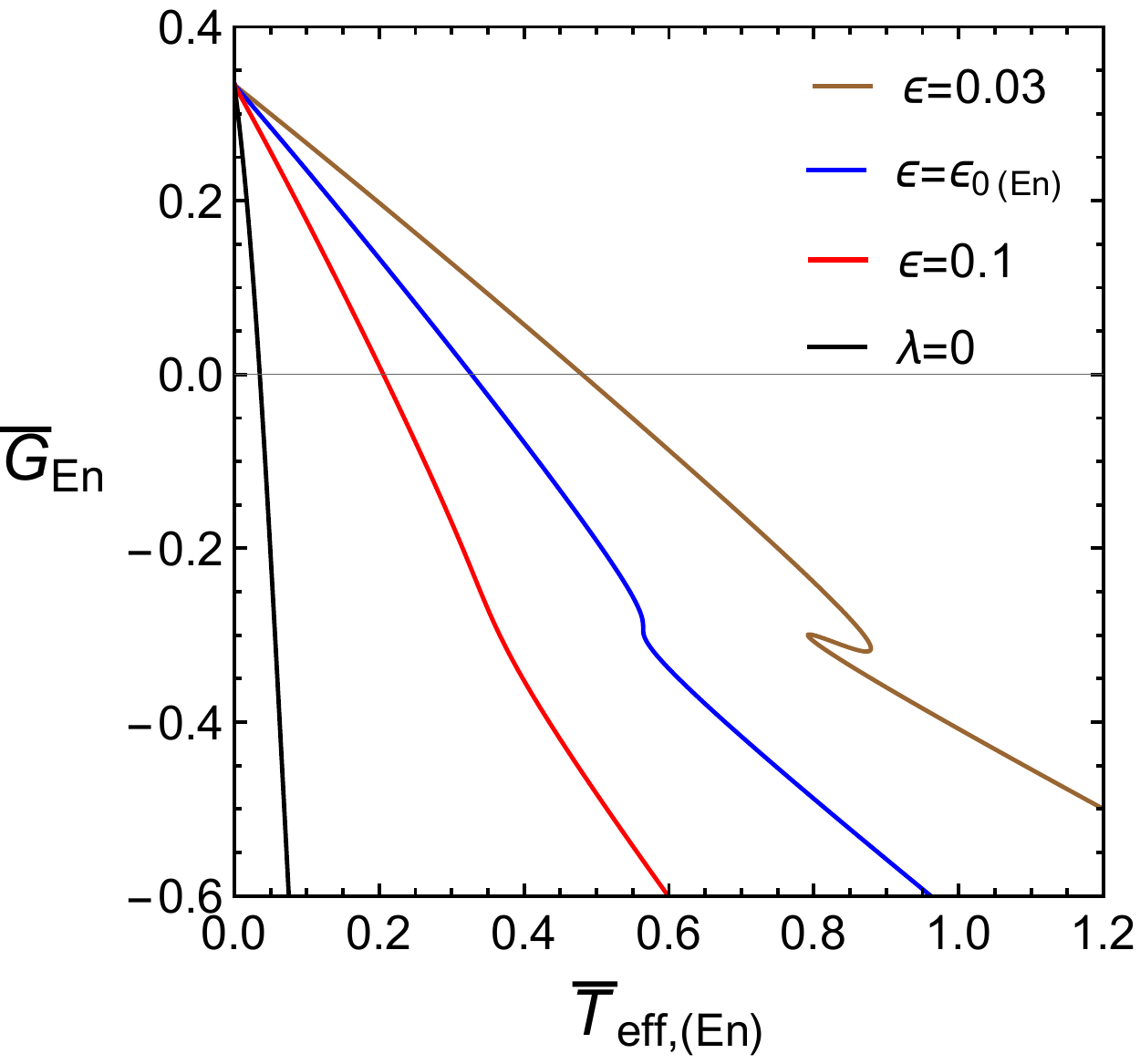}\hspace{1.cm}
\includegraphics[scale=0.45]{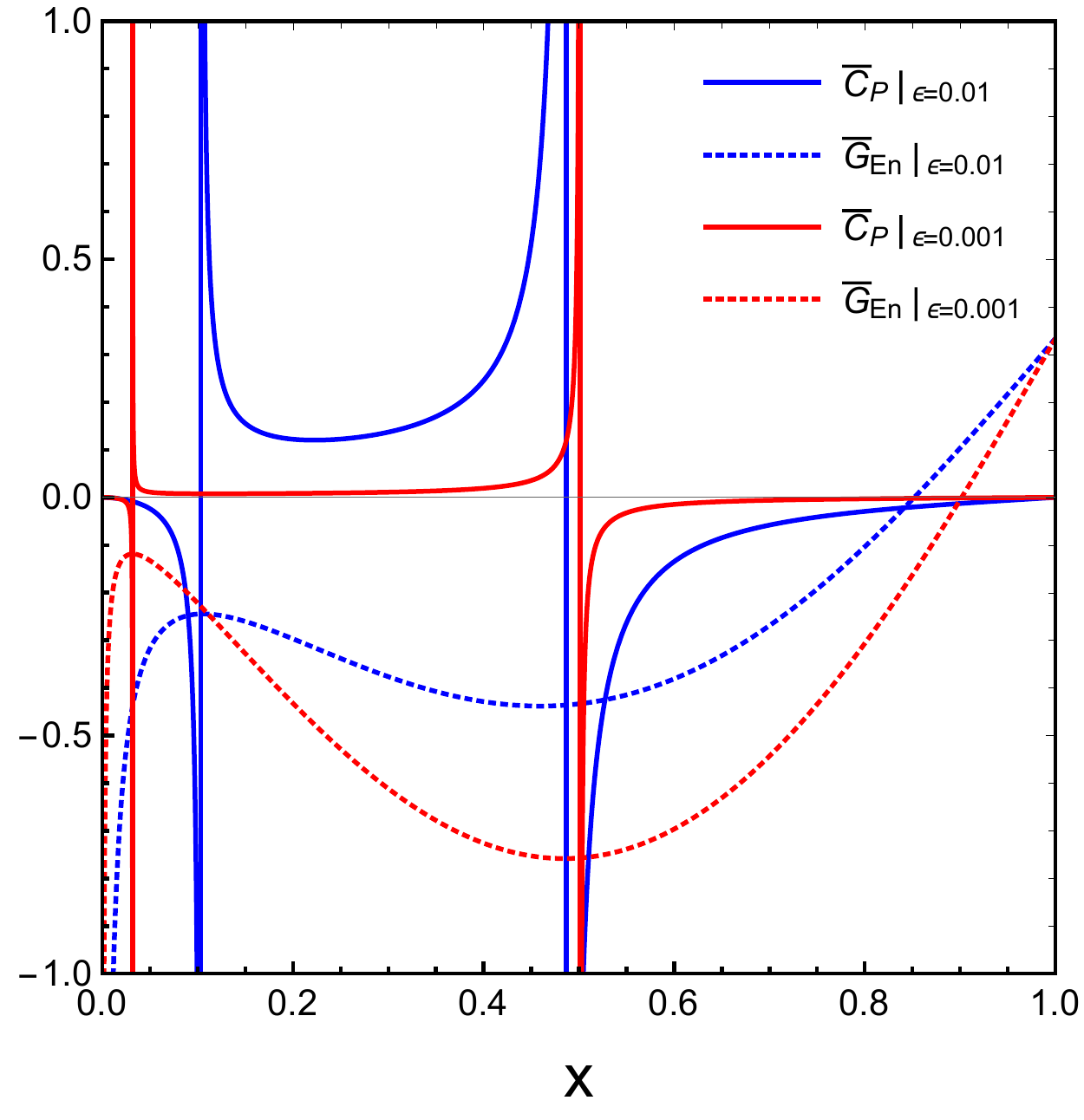}
\end{center}
\caption{The behavior of the Gibbs free energy $G_{\text{En}}$. The left panel shows $G_{\text{En}}$ with various values of $\epsilon$ including the one in the GB limit represented as the black line. \\The right panel shows the comparison between $G_{\text{En}}$ and $C_P$ with $\epsilon=0.01$ (blue lines) and $\epsilon=0.001$ (red lines).}\label{fig:GEn}
\end{figure}

Now let us move our attention to consider the global stability. As we have mentioned in the previous section, the global stability can be determined by considering the value of the Gibbs free energy. According to our effective description, the effective Gibbs free energy can be written as 
\begin{eqnarray}
	\bar{G}_{\text{En}}&=& \sqrt{\Lambda}G_{\text{En}}
	=\sqrt{\Lambda}\left(M-  T_\text{eff,(En)} S\right),\nonumber\\
	&=&\frac{x}{2} \left(1-\frac{x^2}{3}\right) +\frac{\left(1-x^2\right) \left(y^2-1\right) \left(x^2+\epsilon \right) \left(y^2+\epsilon \right) \left[\ln \left(\frac{x^2+\epsilon }{\epsilon }\right)+\ln \left(\frac{y^2+\epsilon }{\epsilon }\right)\right]}{4 (x-y) \big[x y (\epsilon +2)+\epsilon \big]}.\,\,\,\label{GEneff}
\end{eqnarray}
The first part in this equation $\frac{x}{2} \left(1-\frac{x^2}{3}\right)$ is contributed from $M$ which is always positive, implying that the negative part is contributed from another part. Note that it reduces to $\bar{G}_{(b)}$ for $y \rightarrow 0$ and reduces to $\bar{G}_{(c)}$ for $x \rightarrow 0$ similar to the feature of the effective volume. Therefore, three of them reduce to a single value at $x =1$, which is the extremal case. For the behavior of the effective free energy in the left panel in Fig. \ref{fig:GEn}, the free energy is not smooth at the cusps inferred from the diverged points in $C_P$. Since the moderate-sized black hole is locally stable and then provides negative Gibbs free energy, it is globally stable. Note that there is no other bound for $\epsilon$ like for the separated horizon approach, since the free energy for the moderate-sized black hole is always negative. The right panel in Fig. \ref{fig:GEn} also shows that the local extrema of the free energy, $G_\text{En}$, with respect to the variable $x$ are at the divergent points of the heat capacity $C_P$. Moreover, the larger moderate-sized black hole is preferable to form in nature because of the more negative value in the free energy.
\begin{figure}[ht!]
\begin{center}
\includegraphics[scale=0.5]{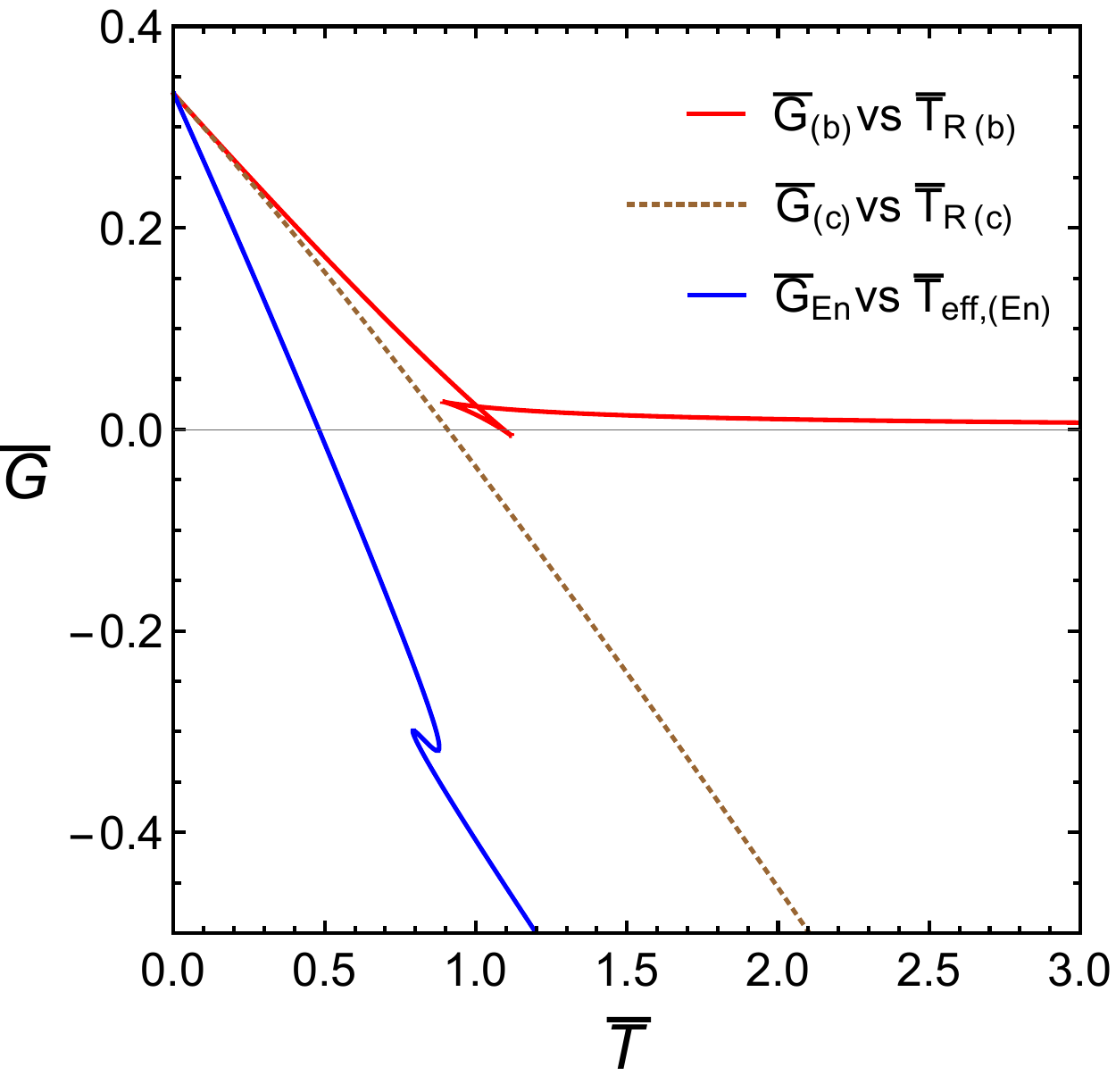}\hspace{1.5cm}
\includegraphics[scale=0.35]{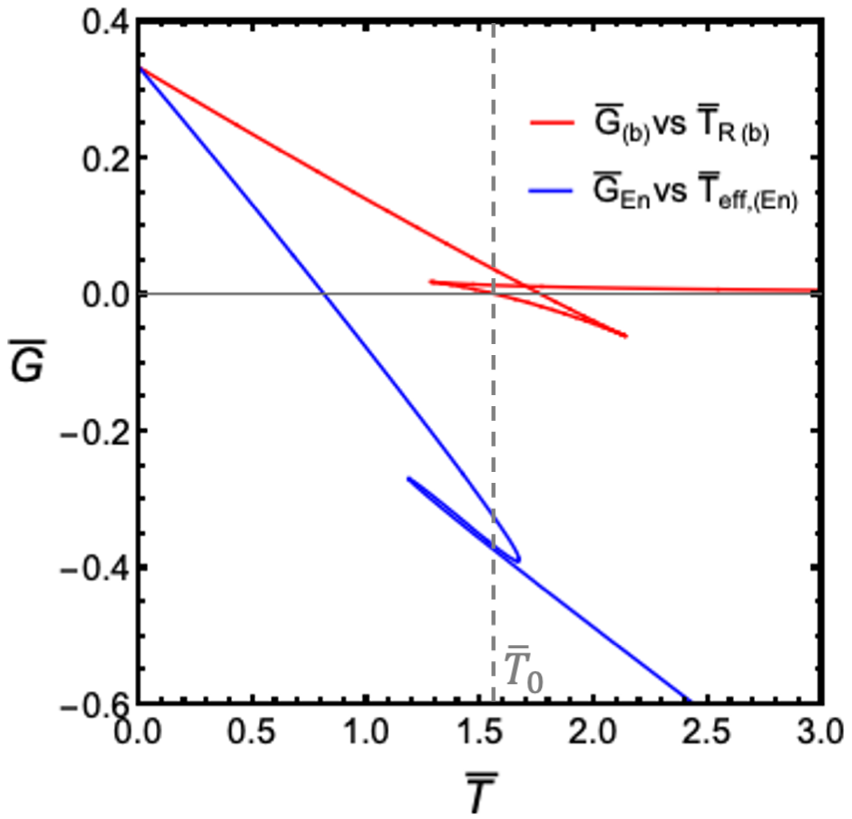}
\end{center}
\caption{Left: the Gibbs free energy $G_{\text{En}}$ compared to the one at the black hole horizon $G_{(b)}$ and the one at the cosmic horizon $G_{(c)}$ versus their own temperatures with $\epsilon = 0.03$. Right: the free energies $G_{\text{En}}$ and $G_{(b)}$ versus their own temperatures with $\epsilon = 0.015$.
}\label{fig:Gall}
\end{figure}

The behaviors of all the free energies can be illustrated in Fig. \ref{fig:Gall}. Also, the effective free energy $G_{\text{En}}$ at a certain temperature is more negative compared to the one for the system at the black hole horizon, $G_{b}$, and the one for the system at the cosmic horizon, $G_{c}$. By comparing the lower bounds in the nonextensive parameter $\lambda$, it is seen that the bound for the separated system is larger than the bound for the effective one, $\lambda_{0G}>\lambda_\text{0(En)}$. The black hole in the effective description can be locally and globally stable in the wider range of $\lambda$.

It is important to note that the hot gas phase has to undergo a zeroth-order phase transition in order to evolve into the moderate-sized stable black hole in the effective system approach. This can be seen from the right panel in Fig.~\ref{fig:Gall} as an example. At the transition temperature denoted by $\bar{T}_0$, there is a transition from the hot gas phase to the black hole phase through which the free energy is discontinuous. While the phase transition at $\bar{T}_0$ is of the first order for the separated system approach, it turns out to be the zeroth-order phase transition at such temperature for the effective system approach. This is the crucial difference between the phase transitions for the effective and separated system approaches.


\subsection{$M$ as internal energy}

The effective system, in which the mass parameter plays the role of the internal energy, is studied in this subsection. The first law for this system should be taken in the form of
\begin{eqnarray}
	\text{d}M=T_\text{eff,(In)}\text{d}S-P_\text{eff}\,\text{d}V,\label{1st law in}
\end{eqnarray}
where $S$ is the total entropy defined in Eq.~\eqref{Seff Renyi} and the volume of this system is assumed as a linear combination between volumes of systems at $r_b$ and $r_c$:
\begin{eqnarray}
	V=V_c+\alpha V_b=\frac{4}{3}\pi\big(r_c^3+\alpha\,r_b^3\big),
\end{eqnarray}
with $\alpha=\pm1$. Similar to the case of $M$ playing the role of enthalpy, the effective quantities, i.e., the temperature $T_\text{eff,(In)}$ and pressure $P_\text{eff}$, can be computed. The effective temperature can be expressed as
\begin{eqnarray}
	T_\text{eff,(In)}
	&=&\Big(\frac{\partial M}{\partial S}\Big)_V
	=\frac{\big(\frac{\partial M}{\partial r_b}\big)_{r_c}\big(\frac{\partial V}{\partial r_c}\big)_{r_b}-\big(\frac{\partial V}{\partial r_b}\big)_{r_c}\big(\frac{\partial M}{\partial r_c}\big)_{r_b}}{\big(\frac{\partial S}{\partial r_b}\big)_{r_c}\big(\frac{\partial V}{\partial r_c}\big)_{r_b}+\big(\frac{\partial V}{\partial r_b}\big)_{r_c}\big(\frac{\partial S}{\partial r_c}\big)_{r_b}},\nonumber\\
	&=&\frac{\left(\pi\lambda r_b^2+1\right)\left(\pi\lambda r_c^2+1\right)\big[-\alpha r_b^5\left(r_b+2r_b r_c\right)+r_c^5\left(r_c+2r_cr_b\right)\big]}{4\pi r_br_c\left(r_b^2+r_br_c+r_c^2\right)^2 \big[\alpha r_b\left(\pi\lambda r_b^2+1\right)+r_c\left(\pi\lambda r_c^2+1\right)\big]}.\label{Teff in}
\end{eqnarray}	
It is seen that this effective temperature is always positively finite for both $\alpha=\pm1$ when $r_b<r_c$, because the terms in square brackets of the numerator and denominator are positive. This result is significantly different from the other effective temperatures in the literature which are not well defined at some value of $r_b/r_c$. In other words, those temperatures can be both positive and negative (their signs are flipped at the divergent points).

To choose the sign of $\alpha$, let us consider the effective temperature at the extremal limit ($r_b=r_c$). The temperature in Eq.~\eqref{Teff in} is infinite for $\alpha=-1$ but vanishes for $\alpha=1$. It is appropriate to choose the constant $\alpha$ as the positive value, since it agrees with $T_\text{R(b,c)}$ or $T_\text{eff,(En)}$ in this limit [see Eqs.~\eqref{TRb}, \eqref{TRc} and \eqref{Teff en}].

\begin{figure}[ht!]
	\begin{center}
	\includegraphics[scale=0.55]{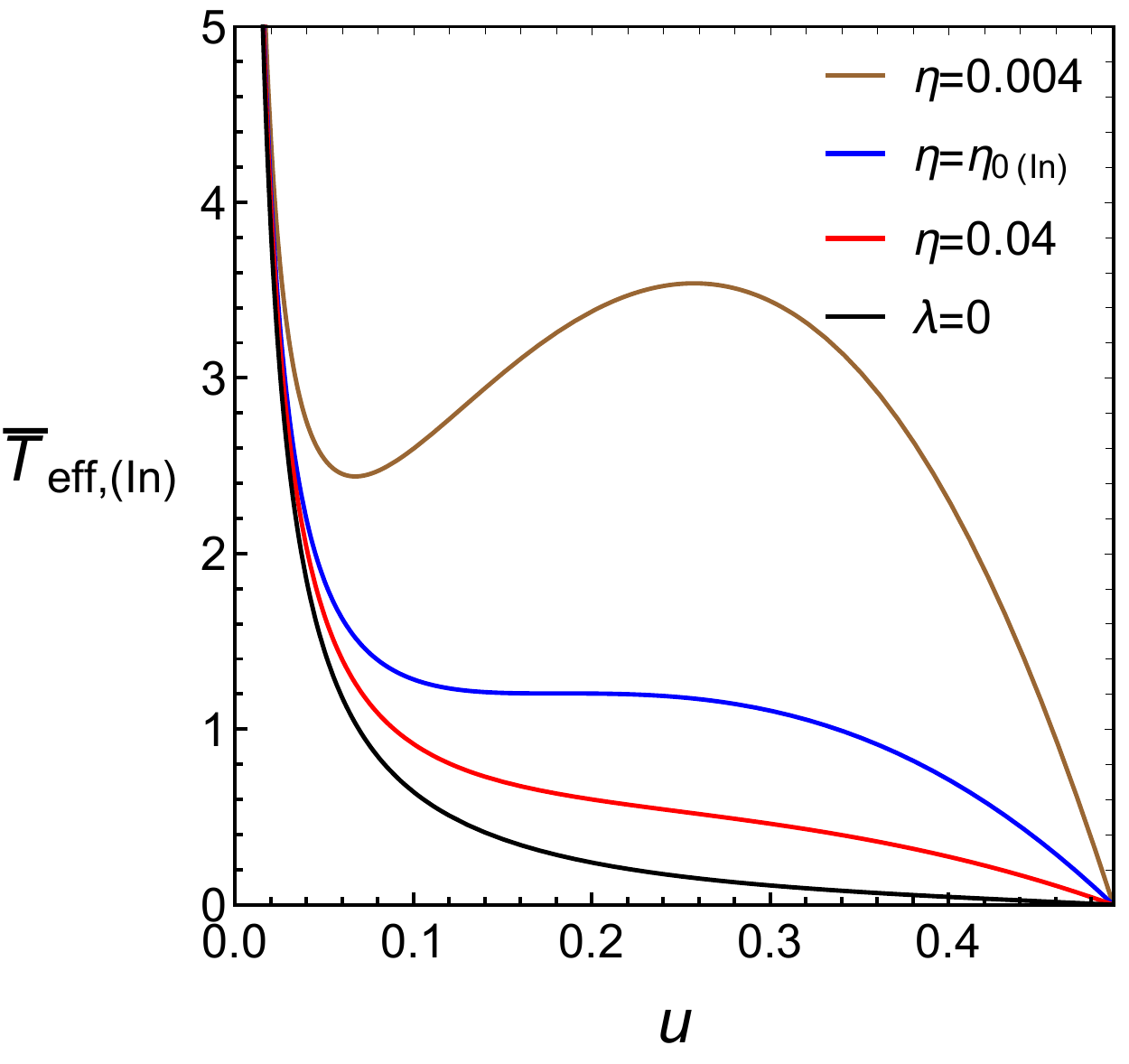}
	\end{center}
	\caption{The (dimensionless) effective temperature $T_\text{eff,(In)}$ versus $u$ for various values of $\eta$ including the one for the GB limit represented as the black line.}\label{fig:Teff-u}
\end{figure}

A difference between effective temperatures is that the temperature $T_\text{eff,(In)}$ cannot be written in terms of $T_{\text{R}(b)}$ and $T_{\text{R}(c)}$, unlike $T_\text{eff,(En)}$ in Eq.~\eqref{re-T}. It is because the temperature $T_\text{eff,(In)}$ is defined under the fixed volume $V$, not the fixed pressure $P$ as the one previously investigated in Sec. \ref{MEn}. To analyze $T_\text{eff,(In)}$ clearly, it is suitable to express the quantities in term of (constant) volume $V$. The dimensionless variables can be defined as follows: 
\begin{eqnarray}
	u=r_b/V^{1/3},\hspace{1cm}
	v=r_c/V^{1/3},\hspace{1cm}
	\eta=1/(\pi\lambda V^{2/3}).
\end{eqnarray}
The variables $u$ and $v$ are related via the dimensionless version of volume, $1=\frac{4}{3}\pi\big(u^3+v^3\big)$. As known that $0<r_b<r_c$, one can find the maximum value of the variable $u$, denoted as $u_\text{max}$, using the fact that the maximum value of $r_b$ is $r_c$ (extremal limit). As a result, one has $u_\text{max}=(\frac{3}{8\pi})^{1/3}\sim0.492$. The effective temperature in Eq.~\eqref{Teff in} is then expressed in terms of the dimensionless variables as 
\begin{eqnarray}
	\bar{T}_\text{eff,(In)}
	=V^{1/3}T_\text{eff,(In)}
	=-\frac{(u^5+u^4v-u^3v^2+u^2v^3-uv^4-v^5)(\eta+u^2)(\eta+v^2)}{4\pi\eta\,uv\big(u^2+uv+v^2\big)^2(\eta+u^2-uv+v^2)},
\end{eqnarray}
which can be written in terms of $u$ (or $v$) and $\eta$. By similar analysis, there exists the multiextremum behavior. The situation in which there exists only one local extremum occurs at $u\sim0.181$ when $\eta=\eta_\text{0(In)}\sim0.0136$. Its profile is shown in Fig.~\ref{fig:Teff-u}.
One can say that the proper upper bound of the parameter $\eta$ is 
\begin{eqnarray}
	\eta<\eta_\text{0(In)}\hspace{.5cm}
	[\text{or }\lambda>\lambda_\text{0(In)}\sim(0.00136\pi V^{2/3})^{-1}],
\end{eqnarray}
because the temperature $T_\text{eff,(In)}$ has positive slope (corresponding to the local stability) when $\eta<\eta_\text{0(In)}$. It is also obvious that there is no local minimum in the GB limit ($\eta\to\infty$).

\begin{figure}[ht!]
	\begin{center}
	\includegraphics[scale=0.55]{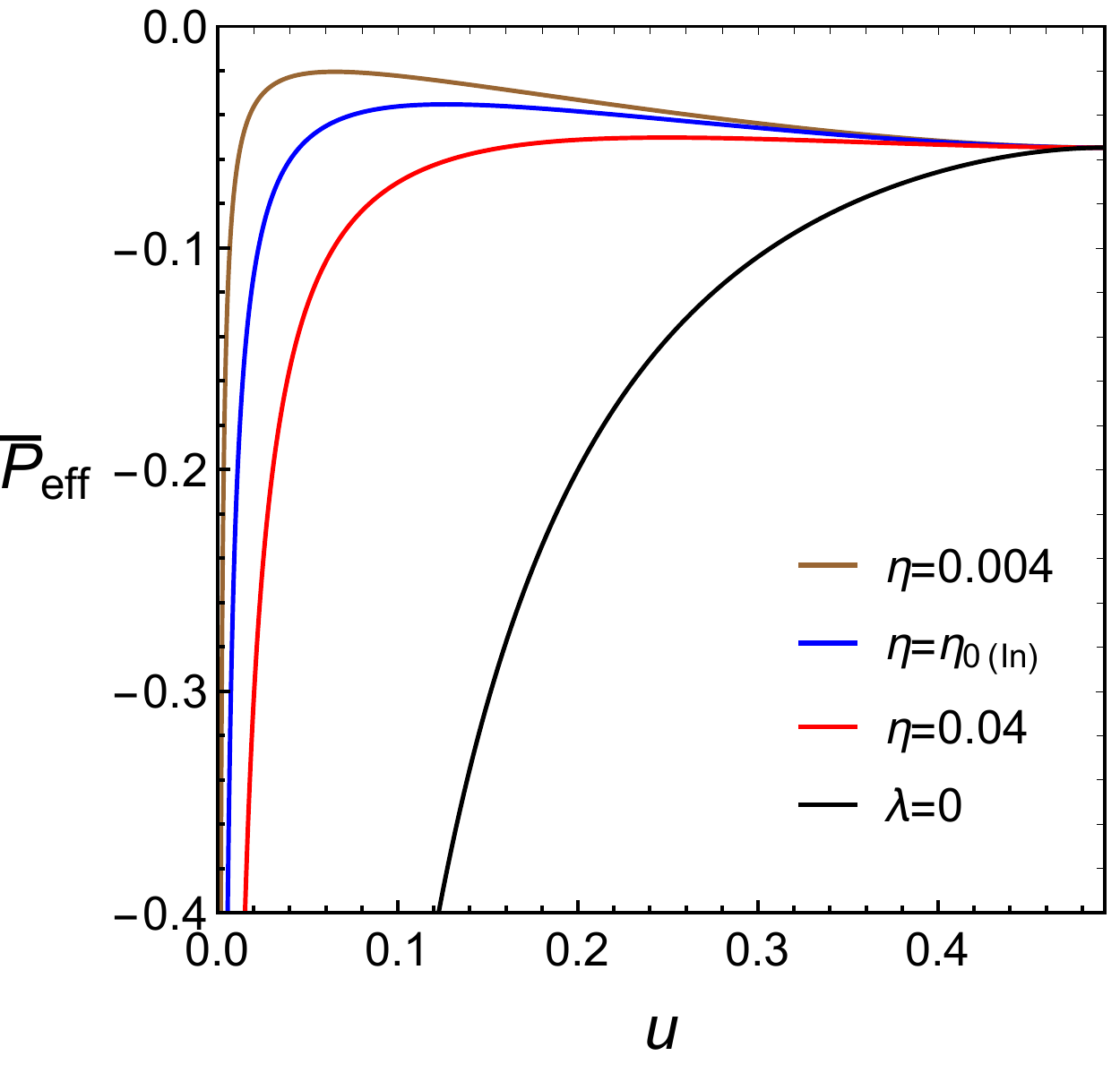}
	\end{center}
	\caption{The effective pressure ($\bar{P}_\text{eff}=V^{2/3}P_\text{eff}$) versus $u$ for various values of $\eta$ including the one for the GB limit represented as the black line.}\label{fig:Peff-u}
\end{figure}

The effective pressure satisfying Eq.~\eqref{1st law in} is obtained as
\begin{eqnarray}	
	P_\text{eff}
	&=&\Big(\frac{\partial M}{\partial V}\Big)_S
	=-\frac{\big(\frac{\partial M}{\partial r_b}\big)_{r_c}\big(\frac{\partial S}{\partial r_c}\big)_{r_b}+\big(\frac{\partial S}{\partial r_b}\big)_{r_c}\big(\frac{\partial M}{\partial r_c}\big)_{r_b}}{\big(\frac{\partial S}{\partial r_b}\big)_{r_c}\big(\frac{\partial V}{\partial r_c}\big)_{r_b}+\big(\frac{\partial V}{\partial r_b}\big)_{r_c}\big(\frac{\partial S}{\partial r_c}\big)_{r_b}}\nonumber\\
	&=&-\frac{r_b^4+r_b^3 r_c-r_b^2 r_c^2+r_b r_c^3+r_c^4+\pi\lambda r_b^2r_c^2(r_b^2+r_b r_c+r_c^2)}{8\pi r_br_c\big(r_b^2+r_b r_c+r_c^2\big)^2\Big[\pi\lambda(r_b^2-r_b r_c+r_c^2)+1\Big]}\nonumber\\
	&=&-\frac{u^2v^2(u^2+uv+v^2)+\eta(u^4+u^3v-u^2v^2+uv^3+v^4)}{8 \pi  u v V^{2/3} \left(u^2+u v+v^2\right)^2 \left(\eta +u^2-u v+v^2\right)}.
\end{eqnarray}
It is obvious that the effective pressure is always negative for the whole range of $r_b\leq r_c$ (or $u\leq v$). In other words, the effective pressure $P_\text{eff}$ still behaves as a tension in this case. Let us consider the properties of this pressure. Similarly to the effective temperature $T_\text{eff,(In)}$, the effective pressure $P_\text{eff}$ can be written in terms of $u$ and $\eta$ when the volume $V$ is fixed. It is found that, in the extremal limit ($u=v=u_\text{max}$), the effective pressure approaches a specific value, $P_\text{eff}(u,\eta)\big|_{u_\text{max}}=-V^{2/3}/(24\pi u_\text{max}^2)\sim-0.055 V^{2/3}$, and its slope $\text{d}P_\text{eff}(u,\eta)/\text{d}u\big|_{u_\text{max}}$ is exactly zero. Notice that these mentioned values at $u=u_\text{max}$ are independent of $\eta$.  On the other hand, the effective pressure goes to negative infinity in the limit $u\to0$, because it is approximated as $-V^{2/3}\eta\Big/\big[6^{1/3}u(4\pi^{2/3}\eta+6^{2/3})\big]$ for very small $u$. Even though the behaviors of $P_\text{eff}(u, \eta)$ at $u=0$ and at $u=u_\text{max}$ do not depend on the parameter $\eta$, the profile of $P_\text{eff}(u, \eta)$ is affected by the existence of $\eta$ in the range of $0<u<u_\text{max}$ as illustrated in Fig.~\ref{fig:Peff-u}.

\begin{figure}[ht!]
	\begin{center}
	\includegraphics[scale=0.55]{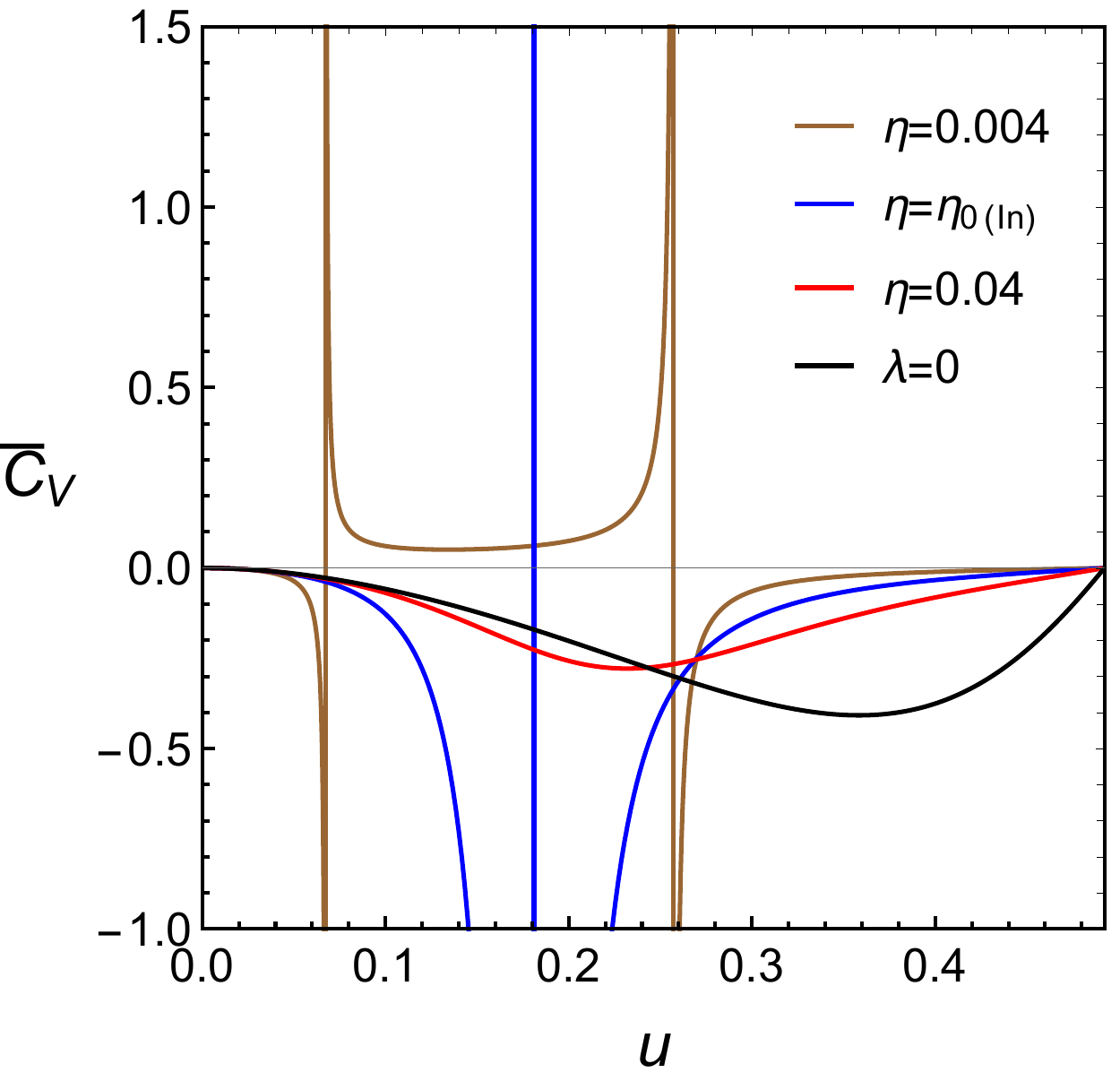}
	\end{center}
	\caption{The (dimensionless) heat capacity $\bar{C}_V=V^{-2/3}C_V$ (solid lines) and temperature $\bar{T}_\text{eff,(In)}$ (dashed lines) versus $u$ for various values of $\eta$ including the one for the GB limit represented as the black line.}\label{fig:CV-u}
\end{figure}

To analyze the local stability of the system, the heat capacity with constant volume is defined via
\begin{eqnarray}
	C_V&=&\left(\frac{\partial M}{\partial T_\text{eff,(In)}}\right)_V=T_\text{eff,(In)}\left(\frac{\partial S}{\partial T_\text{eff,(In)}}\right)_V.
\end{eqnarray}
The expression of $C_V$ is very long and difficult to consider; it is not worth showing here. However, some of its behavior is consequently known from the sign of the slope of the temperature $T_\text{eff,(In)}$. The divergent points of the heat capacity are directly obtained from the points at which slopes of $T_\text{eff,(In)}$ vanish. Hence, there is no divergence in $C_V$ for $\eta>\eta_\text{0(In)}$, while two divergent points appear when $\eta<\eta_\text{0(In)}$. It is easy to check that $M$ is a monotonically increasing function in $u$ for fixed $V$. The sign of $C_V$ is thus the same as that of the slope $\Big(\frac{\partial T_\text{eff,(In)}}{\partial u}\Big)_V$. One can conclude that there is no locally stable range of $u$ for $\eta\geq\eta_\text{0(In)}$, but there exists the locally stable range of $u$ for $\eta<\eta_\text{0(In)}$ as shown in Fig.~\ref{fig:CV-u}.
It is obvious that the moderate-sized black hole is locally stable, similar to the case of $M$ being the enthalpy. By increasing or decreasing $u$ with fixed $V$, there exist transitions between the locally stable and unstable phases for the system with $\eta<\eta_\text{0(In)}$. It is also obvious that the system with GB statistics ($\eta\to\infty$) is always locally unstable.

\begin{figure}[ht!]
	\begin{center}
	\includegraphics[scale=0.5]{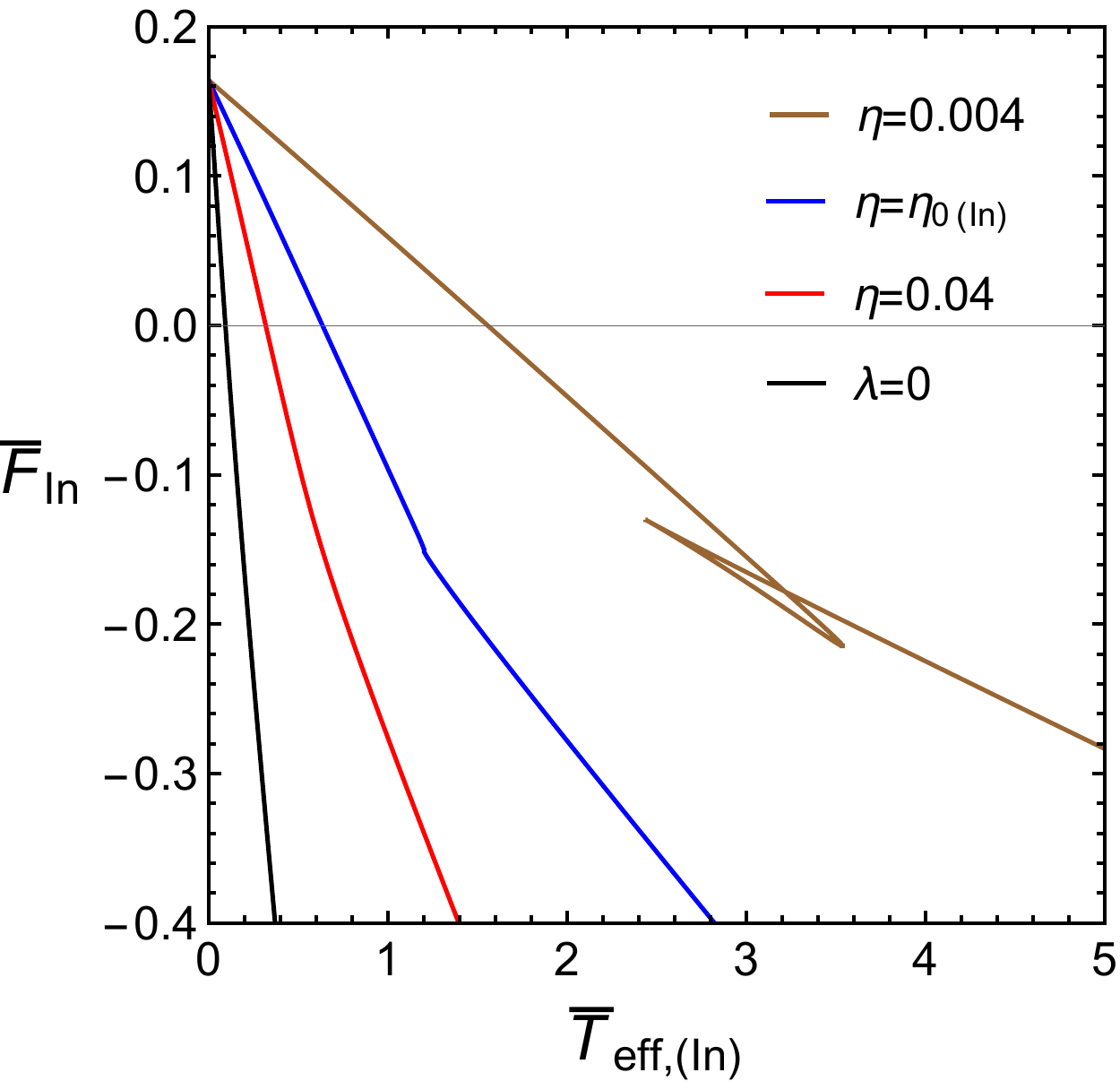}\hspace{1.5cm}
	\includegraphics[scale=0.465]{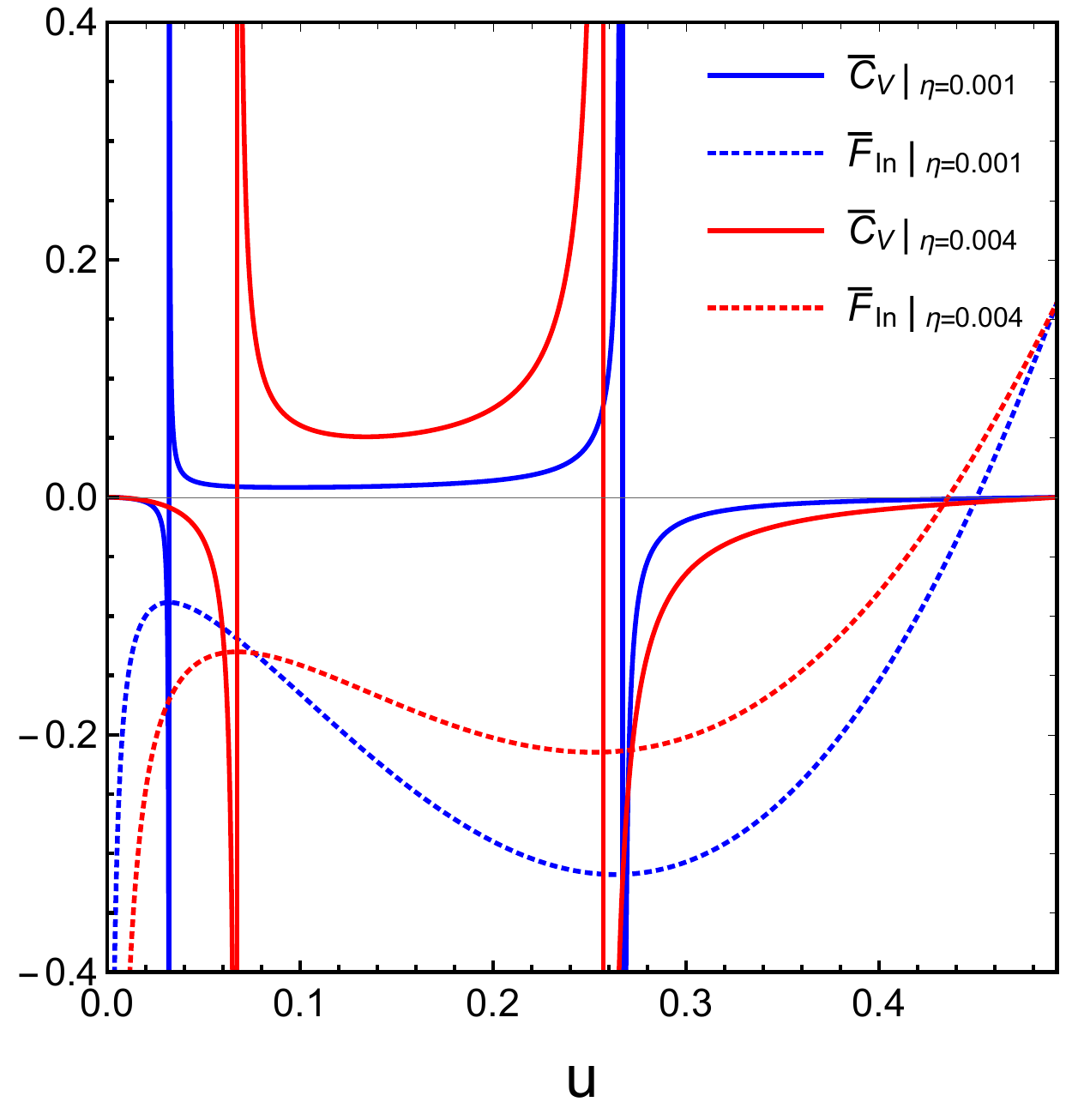}
	\end{center}
	\caption{The behavior of the Helmholtz free energy $F_{\text{In}}$. The left panel shows $F_{\text{In}}$ with various values of $\eta$ including the one in the GB limit represented as the black line. The right panel shows the comparison between $F_{\text{In}}$ and $C_V$ with $\eta=0.001$ (blue lines) and $\eta=0.004$ (red lines).}\label{fig:GIn-T}
\end{figure}

In order to study the global stability of the effective system for the case $M$ being the internal energy, it is more suitable to consider the Helmholtz free energy $F_\text{In}=M-T_\text{eff,(In)}S$ instead of the Gibbs free energy. Since the differentiation of the Helmholtz free energy is $\text{d}F\sim-S\text{d}T_\text{eff,(In)}-P_\text{eff}\text{d}V$, the derivatives of $F_\text{In}$ with respect to $T_\text{eff,(In)}$ for fixed $V$ are proportional to the aforementioned thermodynamical quantities, e.g., $\Big(\frac{\partial F_\text{In}}{\partial T_\text{eff,(In)}}\Big)_V\sim-S$ and $\Big(\frac{\partial^2 F_\text{In}}{\partial T_\text{eff,(In)}^2}\Big)_V \sim C_V$. Hence, the analysis of this free energy is done in the similar way as done previously. The Helmholtz free energy is expressed as
\begin{eqnarray}
	\bar{F}_\text{In}
	&=&V^{-1/3}F_\text{In},\nonumber\\
	&=&\bigg[2 u^2 v^2 \left(u^3+2 u^2 v+2 u v^2+v^3\right) \left(\eta +u^2-u v+v^2\right)\nonumber\\
	&&\hspace{.4cm}+\left(\eta +u^2\right) \left(u^5+u^4 v-u^3 v^2+u^2 v^3-u v^4-v^5\right) \left(\eta +v^2\right) \ln \left(\frac{\eta +u^2}{\eta }\right)\nonumber\\
	&&\hspace{.4cm}+\left(\eta +u^2\right) \left(u^5+u^4 v-u^3 v^2+u^2 v^3-u v^4-v^5\right) \left(\eta +v^2\right) \ln \left(\frac{\eta +v^2}{\eta }\right)\bigg]\nonumber\\
	&&\times\left[4 u v \left(u^2+u v+v^2\right)^2 \left(\eta +u^2-u v+v^2\right)\right]^{-1}.
\end{eqnarray}
Its profiles versus the temperatures $T_\text{eff,(In)}$ are illustrated in the left panel in Fig.~\ref{fig:GIn-T}. It is seen that the swallow tail behavior emerges when $\eta<\eta_\text{0(In)}$. This implies that there exists a phase transition for increasing or reducing $u$ with fixed $V$. It actually corresponds to the transition between the locally stable (moderate-sized) phase and unstable (small- and large-sized) phases. In Fig.~\ref{fig:GIn-T}, the cusp points in the left panel or the extrema in the right panel for the free energy correspond to the divergent points of the heat capacity $C_V$. Moreover, it is not possible to find a further bound of $\eta$ similarly to the case of $M$ being enthalpy, since the free energy $F_\text{In}$ is always negative for the whole range of the moderate-sized black hole. The existence of the nonextensive parameter also identifies that the larger moderate-sized black hole is preferable to form as a locally and global stable system when $\eta<\eta_\text{0(In)}$ (see the right panel in Fig.~\ref{fig:GIn-T}). Similarly, the hot gas phase has to undergo a zeroth-order phase transition in order to evolve into the moderate-sized stable black hole phase in this effective system approach.

As we have discussed through this section, two aspects of the effective systems are constructed. It can be summarized that each approach has a different lower bound in the nonextensive parameter as follows: 
	(i) $\lambda_\text{(En)}\sim\Lambda/(0.0507\pi)$ for $M$ being the enthalpy and 
	(ii) $\lambda_\text{(In)}\sim(0.0136\pi V^{2/3})^{-1}$ for $M$ being the internal energy.
Unfortunately, it may not be possible to argue which approach is better. One just chooses an effective approach which is more appropriate for a given system. One can apply the effective system when $M$ being the enthalpy (internal energy) for the multihorizon black hole with known $\Lambda$ ($V=V_b+V_c$).

Let us emphasize that the effective approach is not the same as studied in the literature. It is modified by introducing the minus sign in front of the change $(\partial S/\partial r_c)_{r_b}$ in the formula for the effective quantities as seen in Appendix~\ref{app: eff quan deriv}. Although we propose the new effective method in order to eliminate the divergence in $T_\text{eff,(En)}$, it fortunately works well in the case $M$ being internal energy. In other words, the temperature $T_\text{eff,(In)}$ has a range with positive slope but has no divergent point. Moreover, for $M$ playing roles of both enthalpy and internal energy, there exists the zeroth-order phase transition from a hot gas phase (with zero free energy) to a moderate-sized stable black hole phase (with negative free energy).


\section{Conclusions}\label{sect: conclusion}

In this work, we have investigated the thermodynamical properties of the Sch-dS black hole using the nonextensive R\'{e}nyi entropy. One of the key signatures of the Sch-dS black hole is the existence of two event horizons called the black hole and cosmic horizons. As a result, there exist two thermodynamical systems with different temperatures. We investigate the thermodynamical properties of the black hole by using both separated system and effective system approaches. For the separated system approach, either thermodynamical system is assumed to be in quasithermal equilibrium. The horizons are very far away from one another, and their temperatures are not much different. For the effective system approach, the Sch-dS black hole as a multihorizon system can be viewed as an effective system with a single temperature. For both approaches, it has been found to be thermodynamically unstable. In this work, we examine the thermodynamics of the Sch-dS black hole with the presence of nonextensive effect. To achieve this, we consider the system using R\'{e}nyi statistics. We have found that the nonextensivity can provide the thermodynamically stable black hole.

In the separated system approach, we follow the first law of thermodynamics in which the mass parameter of the black hole and the cosmological constant are thought of as the enthalpy and thermodynamical pressure, respectively. The thermodynamical stability can be investigated by considering the heat capacity and the free energy. The system is locally stable if the heat capacity is positive and is globally stable if the free energy is negative. From the first law, the sign of the heat capacity changes at the extrema of the temperature so that it is possible to analyze the local stability by considering the slope of the temperature. Using these criteria, one can find the lower bound on the nonextensive parameter from the conditions for the existence of the extrema of the temperature. It is found that there exist local extrema in the temperature for the system at the black hole horizon when the nonextensive parameter is greater than a specific value, $\lambda>\lambda_{0C}=(7+4\sqrt{3})\Lambda/\pi$. In this range of the nonextensive parameter, the slope of the temperature splits into three regions. The middle region (moderate-sized black hole) has positive slope corresponding to the locally stable black hole, and the leftover ones (small and large black holes) have negative slopes corresponding to the locally unstable black hole. Hence, $\lambda_{0C}$ is the lower bound in nonextensive parameter for obtaining the locally stable black hole. Note that the second-order locally stable-unstable phase transition also occurs when $\lambda>\lambda_{0C}$.

For global stability, a stronger bound $\lambda_{0G}\sim\Lambda/(0.0328\pi)>\lambda_{0C}$ is obtained by requiring that the Gibbs free energy must be negative for the whole range of moderate-sized black hole. It is also found that the larger moderate-sized black hole prefers to form in nature, since the free energy is more negative. In the range $\lambda>\lambda_{0G}$, there always exists the first-order Hawking-Page phase transition which is the transition between the thermal radiation or hot gas phase and the locally stable black hole phase. Moreover, to obtain the stability of the whole system, we also investigate the stability of the system corresponding to the cosmic horizon. As a result, we found that, in the viable range of the black hole system $\lambda>\lambda_{0G}$, the system at the cosmic horizon is both locally and globally stable. Therefore, we can argue that the Sch-dS black hole is thermodynamically stable by considering the R\'enyi entropy instead of using the Gibbs-Boltzmann entropy with the condition on the nonextensive parameter as $\lambda>\lambda_{0G}$.

It is interesting to express the lower bounds of $\lambda$ for local and global stabilities in terms of relevant physical length scales. As suggested in Refs.~\cite{Promsiri:2020jga,Promsiri2021}, we can define the so-called nonextensivity length $L_\lambda \equiv 1/\sqrt{\pi \lambda}$, which may be a characteristic length existing in the Sch-dS spacetime.   By comparing with the de Sitter length $L_\Lambda \equiv 1/\sqrt{\Lambda}$, the local stability condition as mentioned above, $\lambda > \lambda_{0C}$, turns out to be in the form of 
\begin{equation}
\frac{L_\lambda}{L_\Lambda} < \Theta_C\qquad \text{(locally stable)}, \label{l_Stable}
\end{equation}
where $\Theta_C=1/\sqrt{7+4\sqrt{3}}\approx 0.268$. In the same way, we can write the global stability condition $\lambda > \lambda_{0G}$ as 
\begin{equation}
	\frac{L_\lambda}{L_\Lambda} < \Theta_G \qquad \text{(globally stable)},\label{g_Stable}
\end{equation}
where $\Theta_G=\sqrt{0.0328}\approx 0.181$.  Note that $\Theta_G < \Theta_C$ means that the global stability condition is stronger than the local stability one, which is consistent with our aforementioned argument.  In the limit $L_\lambda \to \infty$, corresponding to $\lambda \to 0$ (the GB limit), the conditions \eqref{l_Stable} and \eqref{g_Stable} cannot be satisfied.  Namely, this is consistent with that we have no stable Sch-dS spacetime via the GB statistics.  On the contrary, the nonextensive effect becomes important when $L_\Lambda$ is significantly greater than $L_\lambda$, i.e., $L_\lambda$ is very small.  Remarkably, we may think of $1/L_\lambda$ as a fine-graining parameter as suggested in Ref.~\cite{Promsiri2021}.

For the effective approach, it is possible to treat the mass parameter $M$ as both enthalpy and internal energy of the system. In this work, we investigate both issues. The entropy in the effective description is considered as a sum of entropies for the systems at black hole and cosmic horizons which is compatible with the additive composition rule of the R\'enyi entropy. As a result, the effective quantities are derived from the first law for the effective system. Among various kinds of the definition of the effective quantities found in the literature, we propose other suitable forms of the effective quantities in this work. The key idea of this form is to redefine the direction of the heat flow at the cosmic horizon to be opposite to the one at the black hole horizon. The derivation of effective quantities in a general form is provided in Appendix \ref{app: eff quan deriv}. It is worthwhile to note that the definition of the effective quantities defined in this way allows us to avoid the singularity in effective temperature, while the effective temperature defined in the usual way is inevitable to diverge when $T_b=T_c$.

For the effective system in which the mass parameter plays the role of enthalpy, the effective temperature $T_\text{eff,(En)}$ has a region of positive slope or positive heat capacity corresponding to the local stable phase of the system if the nonextensive parameter is large enough similar to the system at the black hole horizon. The lower bound for $\lambda$ in the effective approach is $\lambda_\text{0(En)}\sim\Lambda/(0.0507\pi)$, which is weaker than one for the separated approach, $\lambda_\text{0(En)}<\lambda_{0G}$. Note that the lower bound $\lambda_{0\text{(En)}}$ corresponds to the upper bound of $L_\lambda/L_\Lambda$ with the value $\Theta_\text{En}=\sqrt{0.0507}=0.225$, which is more than $\Theta_G$. This implies that the thermodynamical stability of the black hole in the effective approach requires nonextensive nature of the system less than the one in the separated approach. We also found that there exist particular temperatures in which the black hole in both approaches will be locally stable. In this case, the black hole in the effective approach is always larger than the one in the separated approach. Moreover, there exist particular temperatures for which only black hole in the effective or separated approach is stable. As a result, these particular temperatures can be used to distinguish between the two approaches. For example, if we observe a black hole with such a temperature, it will be argued that the effective or separated approach is the reliable approach.

By considering the Gibbs free energy, we found that there is no further bound on the nonextensive parameter from the Gibbs free energy. This is the one of significant differences between two approaches. In the global stability aspect, the larger moderate-sized black hole is preferred to form, since its free energy is more negative. We found that there is a first-order phase transition between the hot gas and the large unstable black hole. If one wants to realize a transition between the hot gas and the moderate-sized stable black hole, the system must undergo a zeroth-order phase transition between the two prior mentioned phases. This situation is not similar to what happens for the separated system, in which there exists the first-order Hawking-Page phase transition from the the gas to the moderate-sized stable black hole. Moreover, for the hot gas of a specific temperature to evolve into the stable black hole, there is the first-order Hawking-Page phase transition for the separated system, while it must undergo the zeroth-order phase transition for the effective system.

Another possible effective system can be constructed by interpreting $M$ as the internal energy. In this case, the entropy and volume have to be specified. We choose both quantities as the sum of the systems corresponding to black hole and cosmic horizons, $S=S_{\text{R}(b)}+S_{\text{R}(c)}$ and $V=V_b+V_c$. As a result, there exists the lower bound on the nonextensive parameter from the requirement on the existence of the locally stable region, $\lambda_\text{0(In)}\sim1/(0.00136\pi V^{2/3})$. Note that it is not worthy to compare the bounds $\lambda_\text{0(In)}$ and $\lambda_\text{0(En)}$, since these effective systems are defined with different assumptions. The effective system for $M$ being the enthalpy (internal energy) is suitable for describing  the multihorizon black hole with a known cosmological constant (total volume). Although there is no bound from the non-negative Helmholtz free energy, the larger moderate-sized black hole is preferred to form similar to the result in the case of $M$ being enthalpy. Moreover, there exists the zeroth-order phase transition in a similar way as the one in the effective system when $M$ is the enthalpy.

Since the definition of effective quantities presented in this work provides us a way to avoid the singularity in effective temperature for $M$ being both enthalpy and internal energy, it is worthwhile to apply this definition to investigate thermodynamical properties of other black holes with multiple horizons. We leave this investigation for further work. The critical phenomena in black hole thermodynamics is one of the interesting topics which have been intensively investigated recently. Since our work includes the first law of thermodynamics with a specific pressure, it is possible to have a nontrivial equation of state $P=P(T)$. Therefore, such a thermodynamical system may be possible to provide the critical phenomena. This topic is also interesting to investigate in further work.


\section*{Acknowledgements}
We are grateful to Tanapat Deesuwan for helpful discussion. This research project is supported by National Research Council of Thailand (NRCT) : NRCT5-RGJ63009-110. P.W. was supported by the Thailand Research Fund (TRF) through Grant No. MRG6180003 and supported by SERB-DST, India for the ASEAN Project No. IMRC/AISTDF/CRD/2018/000042.


\appendix

\section{Free energy for the system at cosmic horizon}\label{app: free energy at rc}

Let us consider a system in an isothermal process. The heat transfer of this system obeying the second law can be expressed as follows:
\begin{eqnarray}
	\text{d}Q\leq T\text{d}S.
\end{eqnarray}
The equal and less than signs are for the reversible and irreversible processes, respectively. Applying the first law $\text{d}U=\text{d}Q-\text{d}W$, the change of the work done by the system becomes
\begin{eqnarray}
	-\text{d}W\geq\text{d}U-T\text{d}S=\text{d}(U-TS)\equiv\text{d}F,\label{dF}
\end{eqnarray}
which is the differential of the Helmholtz free energy $F=U-TS$. Note that the Helmholtz free energy never increases unless there exists the work done by the isothermal system.

For a system undergoing an isothermal and isobaric transformation, the work term is now expressed as $\text{d}W=P\text{d}V$. Equation~\eqref{dF} is then expressed as
\begin{eqnarray}
	0\geq P\text{d}V+\text{d}F=\text{d}(PV+F)\equiv\text{d}G,
\end{eqnarray}
which is the differentiation of the Gibbs free energy $G=U-TS+PV=F+PV=H-TS$, where $H=U-TS$ is enthalpy. Hence, the Gibbs free energy never increases for a system kept at constant temperature and pressure. In the thermal system at the black hole horizon, the Gibbs free energy can be defined as in Eq.~\eqref{Gbc} ($M$ is treated as the enthalpy in this case).

One now considers the black hole at the cosmic horizon. Recall the first law of this system as
\begin{eqnarray}
	\text{d}M=-T_{\text{R}(c)}\text{d}S_{\text{R}(c)}+V_{c}\text{d}P.
\end{eqnarray}
It implies that the first term on the right-hand side is the heat transfer. This heat term is not possible to have positive value, because the temperature and the change in entropy are not negative. The second law for this system under the isothermal transformation from a state $i$ to a state $f$ is
\begin{eqnarray}
	\int_i^f\frac{-\text{d}Q_{c}}{T_{\text{R}(c)}}\leq S_{\text{R}(c),f}-S_{\text{R}(c),i}.
\end{eqnarray}
Note that the minus sign in front of $\text{d}Q_{c}$ exists because the term $\text{d}Q_{c}$ is always nonpositive. For a constant temperature, one can write
\begin{eqnarray}
	T_{\text{R}(c)}\text{d}S_{\text{R}(c)}\geq-\text{d}Q_{c}
\end{eqnarray}
Applying the first law for the isothermal and isobaric process, $\text{d}U_{c}=\text{d}Q_{c}-P\text{d}V_{c}$, the differentiation of the Gibbs free energy for this system is expressed as
\begin{eqnarray}
	0\geq-\text{d}U_{c}-P\text{d}V_{c}-T_{\text{R}(c)}\text{d}S_{\text{R}(c)}
	&=&\text{d}\big(-U_{c}-PV_{c}-T_{\text{R}(c)}S_{\text{R}(c)}\big)\nonumber\\
	&=&\text{d}\big(-H_c-T_{\text{R}(c)}S_{\text{R}(c)}\big)
	\equiv\text{d}G_{(c)}.
\end{eqnarray}
It is seen that this free energy, $G_{(c)}=-H_c-T_{\text{R}(c)}S_{\text{R}(c)}$, never increases. One chooses to interpret the mass parameter $M$ as a negative value of enthalpy $H_c$. So, the definition of free energy becomes the expression shown in Eq.~\eqref{Gbc}. This also corresponds to the definition of heat capacity in Eq.~\eqref{CPc}:
\begin{eqnarray}
	C_{P(c)}=\bigg(\frac{\partial H_c}{\partial T_{\text{R}(c)}}\bigg)_P
	=-\bigg(\frac{\partial M}{\partial T_{\text{R}(c)}}\bigg)_P.\nonumber
\end{eqnarray}


\section{Effective quantity derivation}\label{app: eff quan deriv}
In this part, the derivation of the effective quantities of the effective systems (in Sec.\ref{sect: eff sys}) is presented. Assume that the differentiation of a state function $M(S,B)$ can be written as
\begin{eqnarray}
	\text{d}M=T_\text{eff}\,\text{d}S+A_\text{eff}\,\text{d}B,\label{dM}
\end{eqnarray}
where $T_\text{eff}$ is the effective temperature and $A_\text{eff}$ is some effective quantity (in this study, it will be the effective pressure or volume). $S$ and $B$ are the given entropy and another state function (in this study, it will be the pressure or volume), respectively, which depend on two horizons $r_b$ and $r_c$:
\begin{eqnarray}
	S=S(r_b,r_c),\hspace{1cm}B=B(r_b,r_c).
\end{eqnarray}
The total derivatives of the above two quantities, thus, are  
\begin{eqnarray}
	\text{d}S&=&\Big(\frac{\partial S}{\partial r_b}\Big)_{r_c}\text{d}r_b-\Big(\frac{\partial S}{\partial r_c}\Big)_{r_b}\text{d}r_c,\label{dS}\\
	\text{d}B&=&\Big(\frac{\partial B}{\partial r_b}\Big)_{r_c}\text{d}r_b+\Big(\frac{\partial B}{\partial r_c}\Big)_{r_b}\text{d}r_c.\label{dP}
\end{eqnarray}
It is important to note that the minus sign in front of the second term on the right-hand side in Eq.~(\ref{dS}) is a consequence of the first law for the cosmic horizon system, $\text{d}M=-T_c\,\text{d}S_c+V_{c}\,\text{d}P$, which has a minus sign in front of the heat term. Let us emphasize that Eq.~(\ref{dS}) is indeed equivalent to Eq.~\eqref{heat_S_minus} in Sec. \ref{sect: intro}. If we consider that the second term on the right-hand side in Eq. (\ref{dS}) corresponds to the energy carried from inside the horizon to the outside one, the entropy of the system increases for the cosmic horizon case, while it decreases for the black hole horizon case. This is due to the fact that the observers are inside the cosmic horizon but outside the black hole horizon. In other words, the observers will experience the positive (negative) energy from the black hole (cosmic) horizon. Therefore, the second term in the right-hand side in Eq. (\ref{dS}) takes the opposite sign to one in the first term. As a result, the temperature at the cosmic horizon, which is positive, can be defined by $T_c=-\frac{\partial M}{\partial S_c}|_P$. In order to take into account the positive temperature at the cosmic horizon to be influenced in the effective temperature, we adopt the argument such that the change of entropy with respect to the cosmic horizon  is in the opposite way to the change with respect to the black hole horizon. The expression of $\text{d}r_b$ can be obtained by considering $\big(\frac{\partial B}{\partial r_c}\big)_{r_b}$[Eq.~\eqref{dS}]$+\big(\frac{\partial S}{\partial r_c}\big)_{r_b}$[Eq.~\eqref{dP}]:
\begin{eqnarray}
	\text{d}r_b
	=\frac{1}{\big(\frac{\partial S}{\partial r_b}\big)_{r_c}\big(\frac{\partial B}{\partial r_c}\big)_{r_b}+\big(\frac{\partial B}{\partial r_b}\big)_{r_c}\big(\frac{\partial S}{\partial r_c}\big)_{r_b}}
	\left[\Big(\frac{\partial B}{\partial r_c}\Big)_{r_b}\text{d}S+\Big(\frac{\partial S}{\partial r_c}\Big)_{r_b}\text{d}B\right].\label{drb}
\end{eqnarray}
Similarly, the expression of $\text{d}r_c$ obtained from $\big(\frac{\partial B}{\partial r_b}\big)_{r_c}$[Eq.~\eqref{dS}]$-\big(\frac{\partial S}{\partial r_b}\big)_{r_c}$[Eq.~\eqref{dP}] is
\begin{eqnarray}
	\text{d}r_c
	=-\frac{1}{\big(\frac{\partial S}{\partial r_b}\big)_{r_c}\big(\frac{\partial B}{\partial r_c}\big)_{r_b}+\big(\frac{\partial B}{\partial r_b}\big)_{r_c}\big(\frac{\partial S}{\partial r_c}\big)_{r_b}}
	\left[\Big(\frac{\partial B}{\partial r_b}\Big)_{r_c}\text{d}S+\Big(\frac{\partial S}{\partial r_b}\Big)_{r_c}\text{d}B\right].\label{drc}
\end{eqnarray}
Since the state function $M$ is also a function of $r_b$ and $r_c$, one can write
\begin{eqnarray}
	\text{d}M&=&\Big(\frac{\partial M}{\partial r_b}\Big)_{r_c}\text{d}r_b+\Big(\frac{\partial M}{\partial r_c}\Big)_{r_b}\text{d}r_c.
\end{eqnarray}
Using Eqs.~\eqref{drb} and \eqref{drc}, the above expression becomes
\begin{eqnarray}
	\text{d}M
	=\left[\frac{\big(\frac{\partial M}{\partial r_b}\big)_{r_c}\big(\frac{\partial B}{\partial r_c}\big)_{r_b}-\big(\frac{\partial B}{\partial r_b}\big)_{r_c}\big(\frac{\partial M}{\partial r_c}\big)_{r_b}}{\big(\frac{\partial S}{\partial r_b}\big)_{r_c}\big(\frac{\partial B}{\partial r_c}\big)_{r_b}+\big(\frac{\partial B}{\partial r_b}\big)_{r_c}\big(\frac{\partial S}{\partial r_c}\big)_{r_b}}\right]\text{d}S
	+\left[\frac{\big(\frac{\partial M}{\partial r_b}\big)_{r_c}\big(\frac{\partial S}{\partial r_c}\big)_{r_b}+\big(\frac{\partial S}{\partial r_b}\big)_{r_c}\big(\frac{\partial M}{\partial r_c}\big)_{r_b}}{\big(\frac{\partial S}{\partial r_b}\big)_{r_c}\big(\frac{\partial B}{\partial r_c}\big)_{r_b}+\big(\frac{\partial B}{\partial r_b}\big)_{r_c}\big(\frac{\partial S}{\partial r_c}\big)_{r_b}}\right]\text{d}B.\nonumber\\
\end{eqnarray}
Comparing to Eq.~\eqref{dM}, one obtains
\begin{eqnarray}
	T_\text{eff}
	&=&\Big(\frac{\partial M}{\partial S}\Big)_B
	=\frac{\big(\frac{\partial M}{\partial r_b}\big)_{r_c}\big(\frac{\partial B}{\partial r_c}\big)_{r_b}-\big(\frac{\partial B}{\partial r_b}\big)_{r_c}\big(\frac{\partial M}{\partial r_c}\big)_{r_b}}{\big(\frac{\partial S}{\partial r_b}\big)_{r_c}\big(\frac{\partial B}{\partial r_c}\big)_{r_b}+\big(\frac{\partial B}{\partial r_b}\big)_{r_c}\big(\frac{\partial S}{\partial r_c}\big)_{r_b}},\\
	A_\text{eff}
	&=&\Big(\frac{\partial M}{\partial B}\Big)_S
	=\frac{\big(\frac{\partial M}{\partial r_b}\big)_{r_c}\big(\frac{\partial S}{\partial r_c}\big)_{r_b}+\big(\frac{\partial S}{\partial r_b}\big)_{r_c}\big(\frac{\partial M}{\partial r_c}\big)_{r_b}}{\big(\frac{\partial S}{\partial r_b}\big)_{r_c}\big(\frac{\partial B}{\partial r_c}\big)_{r_b}+\big(\frac{\partial B}{\partial r_b}\big)_{r_c}\big(\frac{\partial S}{\partial r_c}\big)_{r_b}}.
\end{eqnarray}

For the mass parameter playing a role of an enthalpy, the differential $\text{d}M$ becomes
\begin{eqnarray}
	\text{d}M=T_\text{eff,(En)}\,\text{d}S+V_\text{eff}\,\text{d}P.
\end{eqnarray}	
The effective temperature and volume of the system are, respectively,
\begin{eqnarray}
	T_\text{eff,(En)}
	&=&\Big(\frac{\partial M}{\partial S}\Big)_P
	=\frac{\big(\frac{\partial M}{\partial r_b}\big)_{r_c}\big(\frac{\partial P}{\partial r_c}\big)_{r_b}-\big(\frac{\partial P}{\partial r_b}\big)_{r_c}\big(\frac{\partial M}{\partial r_c}\big)_{r_b}}{\big(\frac{\partial S}{\partial r_b}\big)_{r_c}\big(\frac{\partial P}{\partial r_c}\big)_{r_b}+\big(\frac{\partial P}{\partial r_b}\big)_{r_c}\big(\frac{\partial S}{\partial r_c}\big)_{r_b}},\\
	V_\text{eff}
	&=&\Big(\frac{\partial M}{\partial P}\Big)_S
	=\frac{\big(\frac{\partial M}{\partial r_b}\big)_{r_c}\big(\frac{\partial S}{\partial r_c}\big)_{r_b}+\big(\frac{\partial S}{\partial r_b}\big)_{r_c}\big(\frac{\partial M}{\partial r_c}\big)_{r_b}}{\big(\frac{\partial S}{\partial r_b}\big)_{r_c}\big(\frac{\partial P}{\partial r_c}\big)_{r_b}+\big(\frac{\partial P}{\partial r_b}\big)_{r_c}\big(\frac{\partial S}{\partial r_c}\big)_{r_b}}.
\end{eqnarray}

In the case $M$ being an internal energy, the differentiation of $M$ is the first law of thermodynamics:
\begin{eqnarray}
	\text{d}M=T_\text{eff,(In)}\,\text{d}S-P_\text{eff}\,\text{d}V.
\end{eqnarray}
The effective temperature and pressure of the system are, respectively,	
\begin{eqnarray}
	T_\text{eff,(In)}
	&=&\Big(\frac{\partial M}{\partial S}\Big)_V
	=\frac{\big(\frac{\partial M}{\partial r_b}\big)_{r_c}\big(\frac{\partial V}{\partial r_c}\big)_{r_b}-\big(\frac{\partial V}{\partial r_b}\big)_{r_c}\big(\frac{\partial M}{\partial r_c}\big)_{r_b}}{\big(\frac{\partial S}{\partial r_b}\big)_{r_c}\big(\frac{\partial V}{\partial r_c}\big)_{r_b}+\big(\frac{\partial V}{\partial r_b}\big)_{r_c}\big(\frac{\partial S}{\partial r_c}\big)_{r_b}},\\
	P_\text{eff}
	&=&\Big(\frac{\partial M}{\partial V}\Big)_S
	=-\frac{\big(\frac{\partial M}{\partial r_b}\big)_{r_c}\big(\frac{\partial S}{\partial r_c}\big)_{r_b}+\big(\frac{\partial S}{\partial r_b}\big)_{r_c}\big(\frac{\partial M}{\partial r_c}\big)_{r_b}}{\big(\frac{\partial S}{\partial r_b}\big)_{r_c}\big(\frac{\partial V}{\partial r_c}\big)_{r_b}+\big(\frac{\partial V}{\partial r_b}\big)_{r_c}\big(\frac{\partial S}{\partial r_c}\big)_{r_b}}.
\end{eqnarray}




\end{document}